\documentclass[%
 amsmath,amssymb,
 aps,
 nofootinbib,
]{revtex4}
\usepackage{comment}
\usepackage{xcolor}
\usepackage{physics}
\usepackage{dsfont}
\usepackage{mathtools,slashed}
\usepackage{graphicx}
\usepackage{dcolumn}
\usepackage{bm}
\usepackage{hyperref}

\def\slash#1{\setbox0=\hbox{$#1$}  
   \dimen0=\wd0     
   \setbox1=\hbox{/} \dimen1=\wd1  
   \ifdim\dimen0>\dimen1   
      \rlap{\hbox to \dimen0{\hfil/\hfil}} 
      #1     
   \else     
      \rlap{\hbox to \dimen1{\hfil$#1$\hfil}} 
      /      
   \fi}      %

\newcommand{\sumint}{\sum_{X}\hspace{-0.5cm}\int}

\usepackage{amsmath,bm}
\newcommand{\e}{\mathrm{e}}

\newcommand{\be}{\begin{equation}}
\newcommand{\ee}{\end{equation}}
\newcommand{\bea}{\begin{eqnarray}}
\newcommand{\eea}{\end{eqnarray}}
\newcommand{\nn}{\nonumber}

\begin{document}

\preprint{arXiv:XXXXX.XXXXX [hep-ph]}

\title{Fragmentation contributions to transverse nucleon spin observables in semi-inclusive deep-inelastic scattering at NLO}

\author{Diego Scantamburlo}
\email{diego.scantamburlo@student.uni-tuebingen.de}
\author{Marc Schlegel}
 \email{marc.schlegel@uni-tuebingen.de}
\affiliation{
 Institute for Theoretical Physics, University of T\"ubingen, Auf der Morgenstelle 14, D-72076 T\"ubingen, Germany
}

\date{\today}

\begin{abstract}
We study the spin-dependent cross section in semi-inclusive deep-inelastic lepton-nucleon collisions, $\ell + N^\uparrow\to \ell+h+X$. We focus on the cross section that is integrated over the transverse momentum $\bm{P}_{h\perp}$ of the detected unpolarized hadron. We analyze this cross section at large virtuality $Q^2$ of the exchanged virtual photon within the framework of collinear twist-3 factorization in perturbative QCD. The two main transverse spin observables, the single nucleon spin asymmetry (SSA) and the double longitudinal lepton-transverse nucleon spin asymmetry (DSA), both receive contributions from chiral-odd twist-3 three-parton fragmentation functions. We study these fragmentation contributions to Next-To-Leading Order (NLO) in perturbative QCD. We explicitly observe that collinear twist-3 factorization holds for these contributions at the one-loop level. We confront our NLO formulae with HERMES data and provide numerical predictions for EIC kinematics.
\end{abstract}

\maketitle

\section{Introduction\label{sec:intro}}
The precise understanding of the origin of transverse spin observables in inclusive hadronic processes at high energies within the framework of perturbative QCD (pQCD) has been a longstanding challenge in theoretical hadron physics. Ever since the beginning of the 1980s, surprisingly large transverse polarization effects of $\Lambda$ hyperons have been observed in several experiments in nucleon-nucleon or nucleon-nucleus collisions at Argonne, FermiLab, and CERN \cite{Bunce:1976yb,Schachinger:1978qs,Heller:1978ty,Heller:1983ia,Lundberg:1989hw,Yuldashev:1990az,Ramberg:1994tk,Fanti:1998px}. Somewhat later, in the early 2000s, when polarized proton beams at the Relativistic Heavy Ion Collider (RHIC) became available, 
large transverse single-spin asymmetries (SSA) have also been observed for single-inclusive hadron or jet production in polarized proton collisions $pp^\uparrow \to h/\mathrm{jet}\, X$ ($X$ represents any unobserved hadron in the final state) \cite{Allgower:2002qi,Adams:2003fx,Adler:2005in,Lee:2007zzh,Abelev:2008af,Arsene:2008mi,Adamczyk:2012qj,Adare:2013ekj,Adare:2014qzo,STAR:2020nnl,Adamczyk:2012xd,Bland:2013pkt,PHENIX:2021irw}. 

At the same time, again in the early 2000s, the discovery of certain non-zero transverse single-spin asymmetries in the high-energy process of semi-inclusive hadron production in polarized deep-inelastic lepton-nucleon collisions (SIDIS), $\ell N^\uparrow\to \ell h X$,  the so-called \textit{Sivers effect} \cite{Sivers:1989cc,Sivers:1990fh} and \textit{Collins effect} \cite{Collins:1992kk}, at fixed polarized target experiments like HERMES \cite{Airapetian:1999tv,Airapetian:2001eg,Airapetian:2002mf,Airapetian:2004tw,HERMES:2020ifk} and COMPASS \cite{Alexakhin:2005iw} attracted an enormous amount of interest. This remains true to the present day and will also hold for the future. In fact, a large part of the experimental program at a future Electron-Ion Collider (EIC) is devoted to the experimental measurement of polarized SIDIS \cite{Accardi:2012qut,AbdulKhalek:2021gbh,AbdulKhalek:2022hcn,Burkert:2022hjz,Abir:2023fpo}.

From the perspective of QCD theory, there are two types of approaches to investigate the origin of transverse SSA. One approach is called \textit{collinear twist-3 factorization}. It is based on the collinear factorization approach, which is commonly used in hadronic and particle physics for the pQCD analysis of unpolarized cross sections in terms of collinear parton distribution functions (PDF) and/or fragmentation functions (FF). However, for a power-suppressed observable like a transverse SSA, one generally needs to include more complicated hadronic matrix elements in the collinear twist-3 approach. These matrix elements are often referred to as \textit{multi-parton correlation functions} and/or \textit{multi-parton fragmentation functions}. The collinear twist-3 factorization approach is particularly suitable for the analysis of transverse SSA in single-inclusive hard processes of the form $AB\to CX$, where one of the particles $A$, $B$, or $C$ is transversely polarized. Originally, it was introduced in single-inclusive photon production in polarized proton collisions \cite{Efremov:1981sh,Efremov:1984ip,Qiu:1991pp,Qiu:1991wg}. Later, however, the collinear twist-3 formalism has been used to analyze a variety of transverse single spin asymmetries (SSA) in all sorts of single-inclusive hard processes, such as hadron production in proton collisions \cite{Qiu:1998ia,Eguchi:2006qz,Kouvaris:2006zy,Eguchi:2006mc,Koike:2002gm,Koike:2006qv,Kang:2008ih,Metz:2012ct,Beppu:2013uda}, inclusive deep-inelastic lepton-nucleon collisions (DIS) \cite{Metz:2006pe,Afanasev:2007ii,Metz:2012ui,Schlegel:2012ve}, single-inclusive hadron or jet production in lepton-nucleon collisions \cite{Kang:2011jw,Gamberg:2014eia,Kanazawa:2015ajw,Rein:2025pwu,Rein:2025qhe}, and electron-positron annihilation \cite{Gamberg:2018fwy}. Due to the complexity of the collinear twist-3 factorization approach, most of the aforementioned works calculated the single-spin asymmetries to LO accuracy only. Only very recently, an NLO calculation was presented for an SSA of a \textit{truly}\footnote{By \textit{truly single-inclusive}, we mean processes whose complexity cannot be reduced through a separation into a leptonic and a hadronic tensor. Such a separation is usually performed in (SI)DIS processes, the Drell-Yan (DY) process, or in electron-positron annihilation.} single-inclusive process for the first time \cite{Rein:2025pwu,Rein:2025qhe}. We do note, however, that the applicability of the collinear twist-3 factorization is not limited to single-inclusive processes. As an example, we mention the transverse SSA in semi-inclusive photon production, $\ell N^\uparrow\to\ell \gamma X$, which is also subject to the collinear twist-3 factorization approach \cite{Albaltan:2019cyc,Harris:2025pui}.

Another method to deal with transverse spin observables is the \textit{Transverse-Momentum Dependent (TMD) factorization approach} (cf. the latest review \cite{Boussarie:2023izj}). In this pQCD-based formalism, the fully differential cross section of semi-inclusive processes like SIDIS $\ell N\to\ell h X$ \cite{Mulders:1995dh,Bacchetta:2006tn}, the Drell-Yan (DY) process $hh\to \ell\bar{\ell}X$ \cite{Arnold:2008kf}, or the annihilation process $\ell\bar{\ell}\to h_1h_2X $ \cite{Pitonyak:2013dsu}, is factorized into TMD parton distributions and TMD fragmentation functions that explicitly depend on intrinsic, non-perturbative transverse parton momenta. While extracting TMD PDFs/FFs from experimental semi-inclusive data provides new insights into the 3-dimensional partonic substructure of hadrons, the TMD factorization approach is limited to a specific kinematical regime where the transverse momentum of the final state is much smaller than the hard scale of the process. 

In this paper, we will mainly focus on the SIDIS process. Since most of the experimental data for polarized SIDIS have been gathered at low-energy fixed target experiments like HERMES, COMPASS, and Jefferson Lab (JLab), with the transverse momentum of the final state hadron, $|\bm{P}_{h\perp}|$, being rather small (at most a few GeV), the TMD formalism has proven to be particularly suitable for analyzing this kind of data and extracting TMD PDFs/ FFs.

On the other hand, one can also directly apply the collinear twist-3 formalism to the polarized SIDIS process. This is possible if the transverse momentum $|\bm{P}_{h\perp}|$ of the detected hadron is large, i.e., of the order of the hard scale of the process (for example, the virtuality of the exchanged photon $Q$), or if the transverse momentum $\bm{P}_{h\perp}$ is integrated out. In the latter case, the $\bm{P}_{h\perp}$-integrated SIDIS cross section is no longer fully differential.

From a theoretical perspective, it is particularly convenient to include certain $\bm{P}_{h\perp}$-dependent weighting factors by hand in the $\bm{P}_{h\perp}$-integrals to find a close connection to the fully differential polarized SIDIS structure functions. For example, the SIDIS structure function $F_{UT,T}^{\sin(\phi_h-\phi_s)}$ $\--$ the one that generates the Sivers effect (notation taken from \cite{Bacchetta:2006tn}) $\--$ can be understood in the collinear twist-3 formalism in terms of a particular \textit{quark-gluon-quark} correlation function, the so-called Qiu-Sterman matrix element \cite{Efremov:1981sh,Efremov:1983eb,Efremov:1984ip,Qiu:1991pp,Qiu:1991wg}, once it is weighted with $\bm{P}_{h\perp}$ and followed by a subsequent $\bm{P}_{h\perp}$-integration. Therefore, weighted asymmetries may be considered the ideal observables to learn about multi-partonic matrix elements. Because of this feature, NLO calculations for weighted asymmetries in SIDIS and DY have been performed in the literature \cite{Vogelsang:2009pj,Kang:2012ns,Dai:2014ala,Chen:2016dnp,Chen:2017lvx,Yoshida:2016tfh}.

From an experimental perspective, the inclusion of a weighting factor may be quite difficult to realize, yet not impossible, as a COMPASS measurement of the weighted Sivers asymmetry in SIDIS has demonstrated \cite{COMPASS:2018ofp}. Nevertheless, it may be worthwhile to analyze an unweighted asymmetry as well; that is, the $\bm{P}_{h\perp}$-integrated SIDIS cross section. An NLO analysis of this observable is the main pursuit of this paper.

The motivation for us to do so is twofold: First, upon $\bm{P}_{h\perp}$-integration, most of the azimuthal modulations of the fully differential SIDIS cross section vanish $\--$, including the Sivers effect and the Collins effect. Both are considered leading twist in TMD factorization \cite{Mulders:1995dh,Bacchetta:2006tn}. The only transversely polarized structure functions that survive the $\bm{P}_{h\perp}$-integration are called $F_{UT}^{\sin\phi_s}$ and $F_{LT}^{\cos\phi_s}$, which are azimuthally independent structure functions proportional to $\sin\phi_s$ or $\cos\phi_s$, where $\phi_s$ describes the transverse orientation of the spin vector of the polarized incident nucleon. From the point of view of TMD factorization, these structure functions are power-suppressed and of subleading twist \cite{Mulders:1995dh,Bacchetta:2006tn}. Therefore, the corresponding behavior of the unweighted asymmetries in the collinear twist-3 approach may potentially differ compared to the weighted asymmetries that are related to leading twist TMD effects. We find that, from the point of view of collinear twist-3 factorization, this difference is only marginal.
The second reason we are particularly interested in the chiral-odd twist-3 fragmentation contribution is that, very recently, a somewhat related unweighted asymmetry in the Drell-Yan (DY) process was investigated in collinear twist-3 factorization at NLO accuracy in Ref.~\cite{Zhang:2025fvt}. Specifically, chiral-even quark-gluon-quark correlations in the polarized nucleon were analyzed in Ref.~\cite{Zhang:2025fvt} rather than chiral-odd ones. The validity of collinear twist-3 factorization at the one-loop level was confirmed in Ref.~\cite{Zhang:2025fvt}.
However, in this paper, we focus on the unweighted spin asymmetry in SIDIS rather than DY; this allows us to confront our theoretical formulae obtained in this paper with experimental HERMES data \cite{HERMES:2020ifk}. Since in the LO formula for the $\bm{P}_{h\perp}$-integrated SIDIS structure function $F_{UT}^{\sin\phi_s}$ twist-3 correlations in the polarized nucleon do not contribute \cite{Bacchetta:2006tn}, we turn our attention to the chiral-odd twist-3 effects in the fragmentation process and calculate the NLO corrections. Similar to the findings of Ref.~\cite{Zhang:2025fvt}, we explicitly observe that collinear twist-3 factorization holds at the one-loop level for the chiral-odd fragmentation sector as well.

Our paper is organized as follows: In Section \ref{sec:frames} we discuss the kinematics and the different frames that we use for our calculations. None of this is new and has been mentioned in the literature before, but we nevertheless carefully discuss the frames in detail because the setup of the collinear twist-3 factorization for the $\bm{P}_{h\perp}$-integrated polarized cross section can be quite subtle. The actual setup is also discussed in Section \ref{sec:Setup} in detail for the same reason. In Section \ref{sec:Analytics} we present our analytic results. In Section \ref{sec:Numerics} we confront our NLO result with experimental HERMES data in a numerical study. We find that it is possible to discriminate between various scenarios for the unknown chiral-odd multi-parton fragmentation functions at NLO. Eventually, we provide an NLO prediction for the SIDIS structure function $F_{UT}^{\sin\phi_s}$ at a future EIC. In Section \ref{sec:Conclusions}, we conclude our paper.

\section{Kinematics \& Frames\label{sec:frames}}

In this section, we briefly introduce and summarize the well-known SIDIS kinematics, mostly following the notation of Ref.~\cite{Bacchetta:2006tn}, and we introduce two frames that are relevant for our calculation: the frame in which the incoming nucleon and the produced hadron are collinear (\textit{collinear frame}) and the \textit{Breit frame}. Typically, the QCD factorization is set up in the collinear frame, while the transverse momentum of the detected hadron, $\bm{P}_{h\perp}$, is defined in the Breit frame. Since we consider $\bm{P}_{h\perp}$-integrated SIDIS cross sections, a Lorentz-transformation from the collinear frame to the Breit frame is required to achieve this integration. For leading twist observables like the unpolarized cross section, such a change of frame is implicitly included and does not play a major role. However, the situation is more subtle for subleading twist observables, and one must carefully distinguish between these frames. For this reason, we discuss those frames in more detail in the following. We mention again that none of this is new and has been discussed in the literature, e.g., Refs.~\cite{Mulders:1995dh,Bacchetta:2004jz,Bacchetta:2006tn}.

Let us start by defining the kinematics of the process under consideration, i.e., the semi-inclusive production of unpolarized hadrons in lepton-nucleon collisions, $e(l)+N^\uparrow(P) \to e(l^\prime)+h(P_h)+X$. We assume a one-photon exchange between a massless lepton ($l^2=l^{\prime 2}\simeq 0$) and a nucleon throughout this work. Hence, the space-like momentum of the virtual photon is labeled as $q^\mu=l^\mu-l'^\mu$, and its virtuality $Q^2=-q^2$. The usual SIDIS kinematical variables are given by
\begin{equation}
        x_B=\frac{Q^2}{2P\cdot q},\qquad y=\frac{P\cdot q}{P\cdot l},\qquad z_h=\frac{P\cdot P_h}{P\cdot q}\,.\label{eq:KinVar}
\end{equation}
In this work, we are interested in SIDIS cross sections that are differential in these variables, i.e., $\frac{\mathrm{d}\sigma}{\mathrm{d}y\,\mathrm{d}x_B\,\mathrm{d}z_h}$. However, in an experimental situation, the center-of-mass energy $s=(l+P)^2=\frac{Q^2}{x_B\,y}+M^2$ is typically kept constant, with $M$ being the nucleon's mass ($P^2=M^2$). For this reason, the variables $x_B$ and $y$ are not independent of each other, provided the virtuality $Q^2$ is assumed to be fixed. If the virtuality $Q$ is much larger than the nucleon mass $M$ or the detected hadron's mass $M_h$, i.e., $Q\gg M \sim M_h$, one considers it as the only \textit{hard scale} of the process, and collinear QCD factorization theorems may be applicable. They form the basis of the pQCD calculation presented in this work. Because of the condition $Q\gg M \sim M_h$, we will also neglect all hadronic masses in our work $M\simeq M_h\simeq 0$. Effectively, the hadronic four momenta $P^\mu$ and $P_h^\mu$ become light-like vectors, $P^2=0$ and $P_h^2=0$.

\subsection{Collinear hadron frame\label{sub:collinear}}
As mentioned above, QCD factorization formulae in SIDIS are usually set up in a specific frame in which the nucleon's momentum $P^\mu$ and the detected hadron's momentum $P_h^\mu$ are collinear in the sense that their spatial components determine the $z$-axis of this frame. More importantly, however, is the notion of the term \textit{transverse} in this frame. In general, it takes two light-cone momenta to establish a transverse subspace by means of a projector; in this case, they are the momenta $P^\mu$ and $P_h^\mu$. The transverse projector reads:
\begin{equation}
g_T^{\mu\nu}=g^{\mu \nu}-\frac{P^\mu\,P_h^\nu+P^\nu\,P_h^\mu}{(P\cdot P_h)}= g^{\mu \nu}-\frac{2\,x_B}{z_h\,Q^2}\left(P^\mu\,P_h^\nu+P^\nu\,P_h^\mu\right)\,,\label{eq:gT}
\end{equation}
where we identified $P\cdot P_h=\frac{z_h\,Q^2}{2\,x_B}$, based on Eq.~\eqref{eq:KinVar}.
The collinear frame is then characterized by the following explicit momenta:
\begin{eqnarray}
P^\mu_{\mathrm{coll}}=\left(\frac{Q}{2x_B},0,0,\frac{Q}{2x_B}\right)\,&, & P^\mu_{h,\mathrm{coll}}= \left(z_h\frac{Q}{2},0,0,-z_h\frac{Q}{2}\right)\,.\label{eq:collframe}
\end{eqnarray}
With these momenta, the transverse projector \eqref{eq:gT} indeed becomes a diagonal matrix $g_T=\mathrm{diag}(0,-1,-1,0)$ that projects out the $x$- and $y$-components of any arbitrary four vector 
\begin{equation}
a^\mu=\left(a^0,a^1,a^2,a^3\right) \Longrightarrow a_T^\mu\equiv g_T^{\mu\nu}a_\nu=(0,a^1,a^2,0).\label{eq:aT}
\end{equation}
In order to distinguish between a transverse four vector $a_T^\mu$, as in \eqref{eq:aT}, and a two-dimensional transverse vector, we introduce the notation $\bm{a}_T\equiv \left(a^1,a^2\right)$, with $a_T^2=-\bm{a}_T^2$.

We may then let the projector \eqref{eq:gT} act on the momentum of the virtual photon $q^\mu$. This allows us to identify the transverse photon momentum $q_T^\mu$,
\begin{equation}
    q^\mu = \frac{1}{z_h}\,P_h^\mu -\left(1-\bm{\chi}^2_T\right)\,x_B\,P^\mu+q_T^\mu \quad\Longrightarrow \quad q^\mu_{\mathrm{coll}}=Q\,\left(\tfrac{1}{2}\,\bm{\chi}_T^2,\,\bm{\chi}_T,\,\tfrac{1}{2}\,\bm{\chi}_T^2-1\right)\,,\label{eq:qT}
\end{equation}
with $\bm{\chi}_T\equiv \bm{q}_T/Q$. The collinear frame is theoretically attractive since it allows for a clear kinematical separation of the motion of partons within the nucleon and the fragmentation/hadronization subprocess. However, since the momentum of the detected hadron is fixed, this frame is impractical for studying the transverse momentum integrated SIDIS cross section.

\subsection{Breit frame\label{sub:Breit}}

One may instead perform a Lorentz-transform $\Lambda^\mu_{\,\,\,\nu}$ to a different frame in which the nucleon's momentum $P^\mu$ remains unchanged, but the virtual photon momentum moves along the new $z$-axis and contains zero energy. Such a frame is called the Breit frame (see, e.g., Ref.~\cite{Koike:2006fn}). It may be thought of as a space-like version of a massive particle's rest frame. On the other hand, the explicit form of the nucleon's momentum remains invariant under the Lorentz-transform $\Lambda^\mu_{\,\,\,\nu}$. In the specific Breit frame that we are going to consider in this paper, the momenta $P^\mu$, $q^\mu$ read
\begin{eqnarray}
P^\mu_{\mathrm{Br}}=\left(\frac{Q}{2x_B},0,0,\frac{Q}{2x_B}\right)\,, &\, q^\mu_{\mathrm{Br}}= \left(0,0,0,-Q\right)\,.\label{eq:Brframe}
\end{eqnarray}
The vector $x_B\,P^\mu_{\mathrm{Br}}+q^\mu_{\mathrm{Br}}$ is a light-cone vector as well, and in the Breit frame, we use this vector along with $x_B\,P^\mu_{\mathrm{Br}}$ to characterize the transverse subspace by means of the projector
\begin{equation}
    g_\perp^{\mu\nu}\equiv g^{\mu\nu}-\frac{x_B\,P^\mu\,(x_B\,P^\nu+q^\nu)+x_B\,P^\nu\,(x_B\,P^\mu+q^\mu)}{x_B\,P\cdot(x_B\,P+q)}=g^{\mu\nu}-\frac{2}{Q^2}\left(\,x_B\,P^\mu\,(x_B\,P^\nu+q^\nu)+x_B\,P^\nu\,(x_B\,P^\mu+q^\mu)\right)\,.\label{eq:gperp}
\end{equation}
Note that, in principle, the projectors $g_T$ in \eqref{eq:gT} and $g_\perp$ in \eqref{eq:gperp} map onto different transverse subspaces, and one has to carefully distinguish between them. Indeed, the projector $g_\perp$ has a similar matrix form to the projector $g_T$, that is, $g_\perp^{\mu\nu}=\mathrm{diag}(0,-1,-1,0)$, but in the Breit frame rather than in the collinear hadron frame. In other words, the identification, 
\begin{equation}
a^\mu=\left(a^0,a^1,a^2,a^3\right) \Longrightarrow a_\perp^\mu\equiv g_\perp^{\mu\nu}a_\nu=(0,a^1,a^2,0),\label{eq:aperp}
\end{equation}
of a transverse vector through $g_\perp$ works best in the Breit frame. Eqs.~\eqref{eq:aT} and \eqref{eq:aperp} seem to be alike; however, they are only valid in their respective frames.

The transverse direction of the detected hadron, $P_{h\perp}^\mu$ in the Breit frame, can then be characterized in the following way:
\begin{equation}
    P_h^\mu = z_h\left(\frac{\bm{P}_{h\perp}^2}{z^2_h Q^2}\right)\,x_B\,P^\mu + z_h\,(x_B\,P^\mu+q^\mu)+P_{h\perp}^\mu\,\Longrightarrow P^\mu_{h,\mathrm{Br}}=z_h\,Q\,\left(\tfrac{1}{2}(1+\bm{\chi}_\perp^2),\bm{\chi}_\perp,\tfrac{1}{2}(-1+\bm{\chi}_\perp^2)\right)\,,\label{eq:PhBr}
\end{equation}
with a similar notation $\bm{\chi}_\perp\equiv\bm{P}_{h\perp}/(z_h\,Q)$ and $\bm{P}_{h\perp}^2=-P_{h\perp}^2$. On the other hand, we can re-arrange the decomposition of $q^\mu$ in \eqref{eq:qT} as follows:
\begin{equation}
P^\mu_{h} = -z_h\,\bm{\chi}_T^2\,x_B\,P^\mu+z_h\,(x_B\,P^\mu+q^\mu)-z_h\,q_T^\mu\,.\label{eq:PhqT} 
\end{equation}
By comparing the decompositions \eqref{eq:PhBr} and \eqref{eq:PhqT} of the four vector $P_h^\mu$, we can read off a frame-independent relation between the transverse four vectors $P^\mu_{h\perp}$ and $q^\mu_T$,
\begin{equation}
    P_{h\perp}^\mu = -z_h\,q_T^\mu - 2\,z_h\,\bm{\chi}_T^2\,x_B\,P^\mu\quad\mathrm{or}\quad q_T^\mu = -\tfrac{1}{z_h}P_{h\perp}^\mu - 2\,\bm{\chi}_\perp^2\,x_B\,P^\mu\,.\label{eq:relPhpqT}
\end{equation}
The additional dependence on $x_BP^\mu$ in \eqref{eq:relPhpqT} must be taken into account. It is, however, irrelevant for the squared vectors, i.e.,
\begin{equation}
    q_T^2=\tfrac{1}{z_h^2}P_{h\perp}^2\quad \mathrm{or}\quad \bm{\chi}_T^2 = \bm{\chi}_\perp^2\,.\label{eq:relchi}
\end{equation}
For further details regarding these two frames and the transformation between them, we refer the interested reader to Appendix \ref{appendix:LT}.

\section{Setup of the collinear twist-3 factorization\label{sec:Setup}}
In this section, we discuss the general pQCD setup of the $\bm{P}_{h\perp}$-integrated SIDIS observables that we are going to consider in this paper. We begin our discussion with the common separation of the fully differential SIDIS cross section into a leptonic and hadronic part, assuming a one-photon exchange between the lepton and the nucleon (see, e.g., references \cite{Mulders:1995dh,Bacchetta:2006tn}). Since we anticipate applying dimensional regularization for our NLO calculation, we formulate this separation in $d=4-2\epsilon$ dimensions,
\begin{equation}
    \mathrm{d}\sigma = \frac{1}{(2\pi)^{d-2}\,2\,s}\left(\frac{4\pi\,\alpha_{\mathrm{e.m.}}^2}{Q^4}\right)\,\left(L_{\mu\nu}(l,l^\prime)\,\frac{\mathrm{d}^{d-1}\vec{l}^\prime}{2\,E^\prime}\right)\,\left(W^{\mu\nu}(P,S,q,P_h)\,\frac{\mathrm{d}^{d-1}\vec{P}_h}{2\,E_h}\right)\,,\label{eq:SIDIScs}
\end{equation}
where $\alpha_{\mathrm{e.m.}}\simeq 1/137$ is the electromagnetic fine structure constant, and $L_{\mu \nu}$ is the usual leptonic tensor that can be calculated in QED as
\begin{equation}
    L_{\mu\nu}(l,l^\prime)= 2\left(l^\mu l^{\prime \nu}+l^\nu l^{\prime \mu}-\tfrac{1}{2}Q^2\,g^{\mu\nu}-i\lambda_\ell \,\varepsilon^{ll^\prime\mu\nu}\right)\equiv L_{\mu\nu}^{\mathrm{sym}}(l,l^\prime)-i\lambda_\ell\,L_{\mu\nu}^{\mathrm{anti}}(l,l^\prime)\,.\label{eq:leptTens}
\end{equation}
Note that the leptonic tensor depends on the incoming and outgoing lepton momenta, $l$ and $l^\prime$, and separates into a symmetric (lepton spin independent) part, $L_{\mu\nu}^{\mathrm{sym}}=L_{\nu\mu}^{\mathrm{sym}}$, and an anti-symmetric (lepton spin dependent) part, $L_{\mu\nu}^{\mathrm{anti}}=-L_{\nu\mu}^{\mathrm{anti}}$, where the incoming lepton's helicity is labeled as $\lambda_\ell$. The sign convention for the totally anti-symmetric Levi-Civita tensor is $\varepsilon^{0123}=+1$, and we use a common short-hand notation $\varepsilon^{ll^\prime\mu\nu}=l_\rho l^\prime_\sigma\,\varepsilon^{\rho\sigma\mu\nu}$.\\

The hadronic tensor $W^{\mu\nu}$ in \eqref{eq:SIDIScs} depends on the virtual momentum of the exchanged photon $q^\mu$, the two hadron momenta $P^\mu$ and $P_h^\mu$, and a spin four vector of the polarized nucleon, $S^\mu$. This spin vector satisfies $P\cdot S=0$ and is normalized to $S^2=-1$. In contrast to the leptonic tensor, due to the non-perturbative nature of the strong interactions, the hadronic tensor cannot be calculated directly in perturbation theory. However, for large virtualities $Q^2\gg \Lambda_\mathrm{QCD}^2$, one may apply QCD (TMD) factorization theorems to the SIDIS cross section \cite{Ji:2004wu,Collins:2011zzd,Boussarie:2023izj}. 

Upon integration over the detected hadron's transverse momentum $\bm{P}_{h\perp}$, the direct information on intrinsic partonic momenta in the nucleon or in the fragmentation process is, however, lost. This means that one can expect a factorization of the integrated SIDIS cross section into a hard partonic cross section on the one hand and collinear parton correlations in the nucleon and the fragmentation process on the other. To be specific, the non-perturbative input for the factorized spin-independent unpolarized $\bm{P}_{h\perp}-$integrated SIDIS cross section at large $Q^2$ is given by the usual collinear parton distribution function (PDF) $f_1^{q/g}(x,\mu)$ and the collinear fragmentation function (FF) $D_1^{q/g}(z,\mu)$.

When studying transverse spin-dependent observables, either the PDF or the FF is to be replaced in a factorized description by higher twist correlation functions. These correlation functions may be categorized as \textit{intrinsic}, \textit{kinematical}, or \textit{dynamical} correlation functions \cite{Kanazawa:2015ajw}. In this paper, we solely focus on twist-3 effects in the fragmentation process. However, we plan to study twist-3 effects within the nucleon in a different paper.

The definition of twist-3 fragmentation matrix elements typically depends not only on a light-cone direction dictated by the momentum $P^\mu_h$ of the detected hadron in the final state, but also on an adjoint second light-cone direction provided by an unphysical light-cone vector $n^\mu$ normalized as $P_h\cdot n=1$ and $n^2=0$. While the particular choice of the light-cone vector $n^\mu$ is irrelevant for leading power observables like the unpolarized cross section, the situation becomes more delicate for transverse spin observables. It has been argued in Ref.~\cite{Kanazawa:2015ajw} that for single-inclusive processes, so-called \textit{Lorentz-Invariance Relations} (LIR) among intrinsic, kinematical, and dynamical matrix elements have to be invoked in order to ensure the independence of the collinear twist-3 factorization approach on a specific choice of the light-cone vector $n^\mu$.

The situation is different for a semi-inclusive process like SIDIS. QCD factorization is typically established right away in a specific frame where the two involved hadrons are collinear. We introduced such a frame for SIDIS in section \ref{sub:collinear}: the collinear hadron frame. Here, the adjoint light-cone direction $n^\mu$ is also provided by the factorization theorem as $n^\mu\sim P^\mu$, the physical momentum of the nucleon. In other words, there is no freedom of choice for the light-cone vector $n^\mu$ in SIDIS. This is a major difference compared to single-inclusive observables.

Below, we will further discuss the various twist-3 fragmentation effects contributing to the transverse-spin dependent SIDIS cross section. We emphasize that we will work out these cross sections in a specific light-cone gauge of the gluonic field $G^{\mu,\alpha}$ ($\alpha$ being a color index in the adjoint representation) with antisymmetric boundary conditions,
\begin{equation}
    n\cdot G^\alpha(x)=0\quad \mathrm{and}\quad G_T^{\mu,\alpha}(x\cdot P_h=+\infty)+G_T^{\mu,\alpha}(x\cdot P_h=-\infty)=0\,.\label{eq:LCgauge}
\end{equation}
This gauge is particularly convenient (in comparison to a covariant gauge such as Feynman gauge) to initiate the collinear twist-3 factorization since it reduces the number of redundant terms to a minimum in such a formula. On the other hand, the light-cone gauge complicates the calculation of the partonic hard cross sections as it translates to complications in the numerator of gluonic propagators and/or polarization sums. While in Feynman gauge the numerator of the gluonic propagator/ polarization sum is simply $-g^{\mu\nu}$, it reads in light-cone gauge:
\begin{equation}
    -g^{\mu\nu}\to d^{\mu\nu}(r,n)=-g^{\mu\nu}+\kappa\, \frac{r^\mu n^\nu+r^\nu n^\mu}{r\cdot n}\,.\label{eq:LCprop}
\end{equation}
The light-cone gauge dependent second term proportional to a parameter $\kappa$ depends on a generic (virtual or real) momentum $r^\mu$ that appears in the gluonic propagator/ polarization sums, as well as on the gauge vector $n^\mu$ characterizing the light-cone gauge \eqref{eq:LCgauge}. The denominator $1/(r\cdot n)$ is to be understood as a principal value prescription due to the boundary conditions in \eqref{eq:LCgauge} \cite{Belitsky:2002sm}. We emphasize that in our final result for the spin observables, the dependence on the parameter $\kappa$ in \eqref{eq:LCprop}, in other words, the gauge dependence, drops out in all our final results, indicating color gauge invariance. This is a crucial check of our calculation. In order to achieve the cancellation of $\kappa$, the application of the QCD equation of motion is absolutely essential. We will discuss the QCD equation of motion relations further below.

\subsection{2-parton correlations\label{sub:2Parton}}

\begin{figure}
    \centering
    \includegraphics[width=0.47\linewidth]{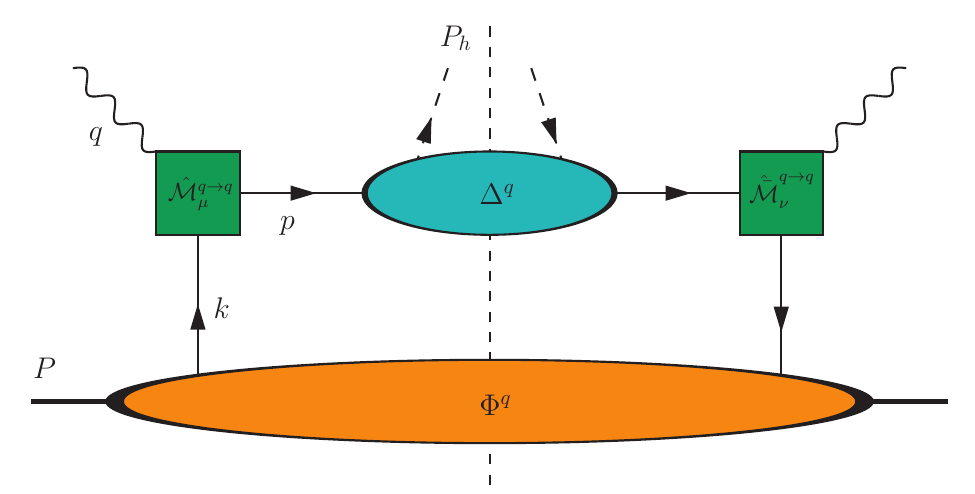}\,\,
     \includegraphics[width=0.47\linewidth]{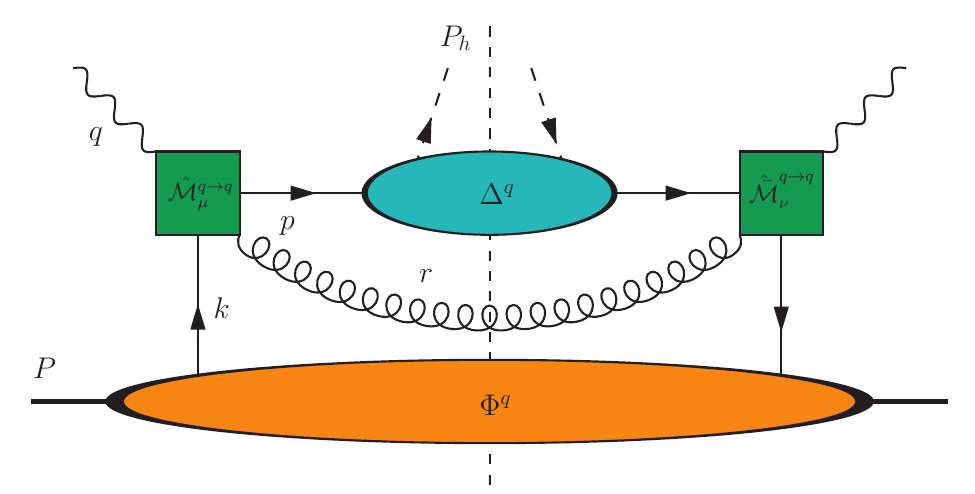}
    \caption{Collinear twist-3 factorization setup: 2-parton fragmentation, \textbf{left:} $n=0$ unobserved partons in the final state, \textbf{right:} $n=1$ unobserved gluon in the final state. 
    }
    \label{fig:2Parton}
\end{figure}

Let us proceed with the collinear twist-3 factorization approach. We first focus on the \textit{intrinsic} and \textit{kinematical} twist-3 contributions. Such contributions are generated by 2-parton (fragmentation) correlators. It is a common approach  \cite{Mulders:1995dh,Bacchetta:2006tn} to choose the following formula as a starting point for the factorization of the hadronic tensor in terms of a non-perturbative (fully unintegrated) parton correlator $\Phi$, a (fully unintegrated) fragmentation correlator $\Delta$, and a partonic scattering amplitude $\mathcal{M}$ (and $\overline{\mathcal{M}}\equiv \gamma^0 \mathcal{M}^\dagger\gamma^0$) that is calculable in perturbation theory,
\begin{equation}
    W^{\mu\nu}_{q\to q}=\int \dd^dk\int\dd^d p\,\sum_{n=0}^\infty\sum_{I_n}\int \dd \mathrm{LIPS}_n\,\sum_qe_q^2\,\mathrm{Tr}\left[\overline{\mathcal{M}}_{q\to q}^\nu\,(k,p,q,\left\{r_n\right\})\,\Delta^q(p)\,\mathcal{M}^\mu_{q\to q}(k,p,q,\left\{r_n\right\})\,\Phi^q(k)\right]\,.\label{eq:q2qstartfact}
\end{equation}
In Eq.~\eqref{eq:q2qstartfact}, we integrate over the full four momenta $k^\mu$ of the quark entering the hard scattering and $p^\mu$ of the fragmenting quark. The additional integration,
\begin{equation}
    \int \dd \mathrm{LIPS}_n = \left\{\begin{matrix}\delta^{(d)}(k+q-p)\,, & n=0 \\ 
    \prod_{i=1}^n\int\tfrac{\dd ^d r_i}{(2\pi)^{d-1}}\,\delta^+(r_i^2)\,\delta^{(d)}(k+q-p-R_n)\,, & n\ge 1\end{matrix}\right.\,,\label{eq:LIPSn}
\end{equation}

indicates a Lorentz-invariant $n$-particle phase space integration in $d$ dimensions over $n$ \textit{unobserved} massless partons produced perturbatively in the hard scattering amplitude, with the set $\{r_n\}$ labeling the four momenta of those partons, and $R_n\equiv\sum_{i=1}^nr_i$. For an NLO calculation, it is sufficient to work with $n=0$ unobserved partons (LO and virtual NLO corrections) and $n=1$ unobserved partons (real NLO corrections). The factorization ansatz is sketched diagrammatically in Fig.~\ref{fig:2Parton}, for $n=0$ and $n=1$. In addition, we have $\delta^{+}(r_i^2)\equiv \Theta(r_i^0)\delta(r_i^2)$ in \eqref{eq:LIPSn}, with $\Theta$ the Heaviside step function. Furthermore, the sum $\sum_{I_n}$ indicates all sorts of summations over the remaining quantum numbers of the unobserved partons, such as color indices, helicities, etc. The term $\mathrm{Tr}$ in \eqref{eq:q2qstartfact} indicates a trace over Dirac and color indices of the correlators and in the amplitudes $\mathcal{M}$, and $e_q$ is the fractional charge of the struck quark.

Now, the factorization ansatz Eq.~\eqref{eq:q2qstartfact} is understood for a fixed momentum $P_h^\mu$ of the detected hadron. It typically serves as a starting point for TMD factorization as well (for $n=0$). Although formulated covariantly, Eq.~\eqref{eq:q2qstartfact} itself is meaningless since it lacks a hierarchy of momentum scales. In order to kinematically single out setups like Eq.~\eqref{eq:q2qstartfact}, we need to invoke such hierarchies through light-cone directions. In other words, we need to expand the parton momenta with respect to these directions. This procedure is called \textit{collinear expansion}. It is particularly convenient to use the light-cone directions defined through the hadron momenta $P$ and $P_h$; therefore, this observation suggests working out the collinear expansion in the collinear hadron frame discussed in section \ref{sub:collinear}.

The collinear expansion of the parton momenta $k$ and $p$ in the collinear momentum frame of section \ref{sub:collinear} is based on the projector $g_T^{\mu\nu}$ in \eqref{eq:gT}. When acting on the four momentum $k^\mu$ of the quark that enters the partonic scattering in the initial state, this projector yields the following decomposition:
\begin{equation}
    k^\mu = \tfrac{2x_B}{z_h Q^2}(k\cdot P_h)\,P^\mu + k_T^\mu + \tfrac{2x_B}{z_h Q^2}(k\cdot P)\,P_h^\mu\,.\label{eq:DefkT}
\end{equation}
The underlying assumption of the collinear expansion is that the quark momentum $k$ is approximately collinear to its parent hadron's momentum $P$. To be specific, the component of $k$ in the direction of $P$ is assumed to be the largest in the hard partonic scattering amplitude $\mathcal{M}$, the transverse component $k_T$ is assumed to be suppressed compared to the component in the direction of $P$, and the virtuality $k^2$ is assumed to vanish. This is the aforementioned momentum hierarchy. Since this paper restricts itself to studying subleading twist contributions in the fragmentation process only, keeping the large component of $k$ and neglecting $k_T$ is already sufficient for the purpose of this work. Therefore, we approximate,
\begin{equation}
\bar{k}^\mu \simeq x\,P^\mu = \tfrac{x_B}{w}\,P^\mu\,,\label{eq:kcollapp}
\end{equation}
with $x\equiv \tfrac{2x_B}{z_h Q^2}(k\cdot P_h) \equiv \frac{x_B}{w}$.
We will use this approximation for $k^\mu$ in the hard scattering amplitude throughout this paper.

The situation is more complicated regarding the momentum of the fragmenting quark. We perform the same decomposition as for $k$, but assume the component of $P_h$ to be the largest; however, we also keep the transverse component, along with the assumption that the fragmenting quark's virtuality is negligible. We therefore approximate in the hard scattering part,
\begin{equation}
    \bar{p}^\mu \simeq \tfrac{1}{z}\,P_h^\mu + p_T^\mu-\tfrac{z}{2}p_T^2\,n^\mu\equiv \tfrac{v}{z_h}\,P_h^\mu + p_T^\mu-\tfrac{z_h}{2\,v}p_T^2\,n^\mu\,,\label{eq:pcollapp}
\end{equation}
with $z\equiv \frac{z_h}{v}$ and the light-cone vector
\begin{equation}
    n^\mu = \frac{2 x_B}{z_h\,Q^2}\,P^\mu\,,\label{eq:LCvector}
\end{equation}
that we also use for the choice of the gauge \eqref{eq:LCgauge}.

We will now apply the collinear approximations \eqref{eq:kcollapp}, \eqref{eq:pcollapp} to the factorization ansatz \eqref{eq:q2qstartfact}, and we obtain the hadronic tensor for $n=0$ unobserved partons,
\begin{equation}
    W^{\mu\nu}_{q\to q,v}\simeq\int \dd x\int\tfrac{\dd z}{z^2}\int \dd^{d-2}\bm{p}_T\,\delta^{(d)}(\bar{k}+q-\bar{p})\sum_qe_q^2\,\mathrm{Tr}\left[\overline{\mathcal{M}}_{q\to q,v}^\nu\,(\bar{k},\bar{p},q)\,\Delta^q(z,\bm{p}_T)\,\mathcal{M}^\mu_{q\to q,v}(\bar{k},\bar{p},q)\,\Phi^q(x)\right]+\mathcal{O}(\tfrac{\Lambda^2}{Q^2})\,,\label{eq:q2qv}
\end{equation}
and for $n=1$ unobserved partons,
\begin{eqnarray}
    W^{\mu\nu}_{q\to q,r}&\simeq&\int \dd x\int\tfrac{\dd z}{z^2}\int \dd^{d-2}\bm{p}_T\,\delta^{+}\left((\bar{k}+q-\bar{p})^2\right)\tfrac{1}{(2\pi)^{d-1}}\sum_qe_q^2\,\sum_{I_1}\times\nn\\
    &&\mathrm{Tr}\left[\overline{\mathcal{M}}_{q\to q,r}^\nu\,(\bar{k},\bar{p},q)\,\Delta^q(z,\bm{p}_T)\,\mathcal{M}^\mu_{q\to q,r}(\bar{k},\bar{p},q)\,\Phi^q(x)\right]+\mathcal{O}(\tfrac{\Lambda^2}{Q^2})\,.\label{eq:q2qr}
\end{eqnarray}
We emphasize that the underlying decomposition of the integration measure into two light-cone integration variables and a $d-2$ dimensional transverse Euclidean integration, $\dd ^d p=\dd (p\cdot n)\,\dd^{d-2}\bm{p}_T\,\dd (p\cdot P_h) $, which is implicit in \eqref{eq:q2qv} and \eqref{eq:q2qr}, is only meaningful in the collinear hadron frame of section \ref{sub:collinear}. Note that Eqs.~\eqref{eq:q2qv}, \eqref{eq:q2qr} include the collinear correlator $\Phi(x)$ and the TMD correlator $\Delta(z,\bm{p}_T)$. But before we provide explicit definitions for these non-perturbative quantities, let us further dwell on the kinematics and frames.
\subsubsection{Virtual contributions, $n=0$:}
Let us first focus on the situation of $n=0$ unobserved final state partons, i.e., Eq.~\eqref{eq:q2qv}. Thanks to the explicit form \eqref{eq:collframe} of the momenta $P$, $P_h$ in the collinear hadron frame, it is possible to cleanly separate the $d$-dimensional delta function in \eqref{eq:q2qv} as
\begin{equation}
    \delta^{(d)}(\bar{k}+q-\bar{p}) = x_Bz_h\,\tfrac{2}{Q^2}\,\delta(x-x_B)\,\delta(z-z_h)\,\delta^{(d-2)}(\bm{q}_T-\bm{p}_T)\,.\label{eq:deltacollframe}
\end{equation}
Using this separation, the hadronic tensor \eqref{eq:q2qv} simplifies as:
\begin{equation}
    W^{\mu\nu}_{q\to q,v}=\int \dd^{d-2}\bm{p}_T\,\delta^{(d-2)}(\bm{q}_T-\bm{p}_T)\,\tfrac{2x_B}{z_hQ^2}\sum_qe_q^2\,\mathrm{Tr}\left[\overline{\mathcal{M}}_{q\to q,v}^\nu\,(\bar{k},\bar{p},q)\,\Delta^q(z_h,\bm{p}_T)\,\mathcal{M}^\mu_{q\to q,v}(\bar{k},\bar{p},q)\,\Phi^q(x_B)\right]^{x=x_B}_{z=z_h}.\label{eq:q2qv2}
\end{equation}
In this formula, the momentum $P_h$ is still fixed in the collinear hadron frame. However, we wish to study the $\bm{P}_{h\perp}$-integrated cross section in the Breit frame. For this reason, we perform a Lorentz-transformation of the hadronic tensor \eqref{eq:q2qv2} to the Breit frame first, as discussed in section \ref{sub:Breit}. The crucial point here is to carefully take into account the transformation of the transverse vector $p_T^\mu$ from the collinear frame to the Breit frame. This can be done by means of the projector $g_\perp^{\mu\nu}$ in Eq.~\eqref{eq:gperp}. In \eqref{eq:relPhpqT}, we have established the relation between the four vector $q^\mu$ and $P_{h\perp}^\mu$. However, if we consider  the Euclidean $d-2$ dimensional transverse vectors $\bm{q}_T$ and $\bm{P}_{h\perp}$, we can identify 
\begin{equation}
\bm{q}_T=-\tfrac{1}{z_h}\bm{P}_{h\perp}\,.\label{eq:relqTPhp}
\end{equation}
One must keep in mind that the left hand side of \eqref{eq:relqTPhp} is to be understood in the collinear hadron frame, while the right hand side is in the Breit frame.

A similar relation can be derived for the four momentum $p_T^\mu$, as well as for the $d-2$ transverse Euclidean vector $\bm{p}_T$,
\begin{eqnarray}
    p_{T,\mathrm{Br}}^\mu = (p_T)_\perp^\mu + 2 x_B\,\tfrac{(\bm{\chi}_\perp)\cdot{(\bm{p}_T)_\perp}}{Q}\,P^\mu, &\quad \mathrm{but}\quad & \bm{p}_{T,\mathrm{Br}}= (\bm{p}_T)_\perp\,.\label{eq:relpT}
\end{eqnarray}
The hadronic tensor \eqref{eq:q2qv2} then reads, after a Lorentz-transform to the Breit frame,
\begin{eqnarray}
    W^{\mu\nu}_{q\to q,v,\mathrm{Br}}&=&\int \dd^{d-2}(\bm{p}_T)_\perp\,\delta^{(d-2)}((\bm{p}_T)_\perp+\tfrac{1}{z_h}\bm{P}_{h\perp})\,\tfrac{2x_B}{z_hQ^2}\times\nn\\
    &&\sum_qe_q^2\,\mathrm{Tr}\left[\overline{\mathcal{M}}_{q\to q,v}^\nu\,(\bar{k}_{\mathrm{Br}},\bar{p}_{\mathrm{Br}},q_{\mathrm{Br}})\,\Delta^q(z_h,(\bm{p}_T)_\perp)\,\mathcal{M}^\mu_{q\to q,v}(\bar{k}_{\mathrm{Br}},\bar{p}_{\mathrm{Br}},q_{\mathrm{Br}})\,\Phi^q(x_B)\right]\,.\label{eq:q2qvBr}
\end{eqnarray}
Here, $\bar{k}^\mu_{\mathrm{Br}}=x_B\,P_{\mathrm{Br}}^\mu$, $P_{\mathrm{Br}}$, and $q_{\mathrm{Br}}$ are given in Eq.~\eqref{eq:Brframe}. Curiously, after applying the condition $(\bm{p}_T)_\perp+\tfrac{1}{z_h}\bm{P}_{h\perp}=0$, which is enforced by the delta function in \eqref{eq:q2qvBr}, the forms of the four vectors $\bar{p}_{\mathrm{Br}}$ and $\bar{p}_{T,\mathrm{Br}}$ simplify significantly in the Breit frame. We find
\begin{equation}
    \bar{p}^\mu_{\mathrm{Br}}=x_B\,P^\mu_{\mathrm{Br}} + q^\mu_{\mathrm{Br}}\,,\label{eq:pBr}
\end{equation}
which indicates that $\bar{p}^\mu_{\mathrm{Br}}$ does neither depend on $(\bm{p}_T)_\perp$ nor on $\bm{P}_{h\perp}$! Consequently, the partonic amplitudes $\mathcal{M}^\mu_{q\to q,v}$ and $\overline{\mathcal{M}}_{q\to q,v}^\nu$ do not depend on the transverse momenta $(\bm{p}_T)_\perp$ and $\bm{P}_{h\perp}$ either, to all orders in perturbation theory.

Another advantage of the Breit frame is that we can easily study the $\bm{P}_{h\perp}$-integrated SIDIS cross section. In particular, the leptonic tensor $L_{\mu\nu}$ in \eqref{eq:leptTens} does not depend on the transverse hadron momentum $\bm{P}_{h\perp}$ in this frame, and neither do the other kinematical prefactors in \eqref{eq:SIDIScs}. Therefore, in the Breit frame, we can directly integrate the hadronic tensor and study collinear twist-3 factorization at the level of the hadronic tensor. In the Breit frame, the Lorentz-invariant phase space measure can be decomposed as
\begin{equation}
    \frac{\dd^{d-1}\vec{P}_h }{2E_h}=\frac{\dd z_h}{2z_h}\dd^{d-2}\bm{P}_{h\perp}\,.\label{eq:measurePhperpBr}
\end{equation}
In the following, we will integrate out the transverse hadron momentum $\bm{P}_{h\perp}$ but keep the variable $z_h$ fixed. Additionally, we keep the differential $\dd z_h$ explicitly, noting that after contraction with the leptonic tensor and insertion into the master formula \eqref{eq:SIDIScs}, the cross section remains differential in the variable $z_h$. With these comments, the $\bm{P}_{h\perp}$-integrated hadronic tensor \eqref{eq:q2qvBr} further simplifies as
\begin{eqnarray}
    \tilde{W}^{\mu\nu}_{q\to q,v}&\equiv& \frac{\dd z_h}{2\,z_h}\int \dd^{d-2}\bm{P}_{h\perp} W^{\mu\nu}_{q\to q,v,\mathrm{Br}} \nn\\
    &=&\dd z_h\, z_h^{-2\epsilon}\,\tfrac{x_B}{Q^2}\times\nn\\
    &&\hspace{-0.5cm}\sum_qe_q^2\,\mathrm{Tr}\left[\overline{\mathcal{M}}_{q\to q,v}^\nu\,(\bar{k}_{\mathrm{Br}},\bar{p}_{\mathrm{Br}},q_{\mathrm{Br}})\,\left(\int \dd^{d-2}(\bm{p}_T)_\perp\,\Delta^q(z_h,(\bm{p}_T)_\perp)\right)_{\mathrm{Br}}\,\mathcal{M}^\mu_{q\to q,v}(\bar{k}_{\mathrm{Br}},\bar{p}_{\mathrm{Br}},q_{\mathrm{Br}})\,\Phi^q(x_B)\right]\,.\label{eq:q2qvInt}
\end{eqnarray}
This is a remarkably simple result.\\

At this point, let us now discuss the (unrenormalized, but regularized in $d$ dimensions) correlators $\Phi$ and $\Delta$. Since we consider twist-3 effects in the fragmentation process only, the correlator $\Phi^q(x)$ is a collinear matrix element that can be parameterized in terms of leading twist collinear parton distribution functions as follows (see, e.g., \cite{Kanazawa:2015ajw})
\begin{eqnarray}
    \Phi_{ij}^q(x) & = & \int_{-\infty}^\infty \tfrac{\dd \lambda}{2\pi}\,\e^{i\lambda x}\langle P,S|\,\bar{q}_j(0)\,[0;\lambda\,m]\,q_i(\lambda\,m)\,|P,S\rangle\nn\\
    &=&\tfrac{1}{2}\,\slash{P}_{ij}\,f_1^q(x)-\tfrac{1}{4}\left([\slash{P},\slash{S}]\gamma_5\right)_{ij}\,h_1^q(x)+...\,.\label{eq:Phi}
\end{eqnarray}
The matrix element in the first line of \eqref{eq:Phi} contains two quark fields $q$ and $\bar{q}$ separated along a light-cone direction $m^\mu$. The Wilson line $[0;\lambda\,m]$ connects the arguments of the quark fields via a straight line and ensures gauge invariance of the matrix element. Recall that the initial version of factorization for the SIDIS hadronic tensor \eqref{eq:q2qstartfact} has been set up in the collinear hadron frame where the hadron momenta $P^\mu$ and $P_h^\mu$ play the role of each other's adjoint light-cone directions. Hence, it is $m^\mu = \frac{2 x_B}{z_h\,Q^2}P_h^\mu$. The second line displays two leading Dirac structures accompanied by the distribution function $f_1^q(x)$ of unpolarized quarks in an unpolarized nucleon and the transversity function $h_1^q(x)$ \cite{Ralston:1979ys,Artru:1989zv,Jaffe:1991kp,Jaffe:1991ra,Cortes:1991ja}. The latter describes the distribution of transversely polarized quarks in a transversely polarized nucleon. Other structures, indicated by $...$, are also present \cite{Kanazawa:2015ajw}, but are not relevant in this work. 

We need to take extra care with the fragmentation correlator in \eqref{eq:q2qvInt}. One might be tempted to work with a collinear fragmentation correlator in \eqref{eq:q2qvInt} directly. However, the TMD fragmentation correlator $\Delta^q(z,\bm{p}_T)$ originally appeared in the factorization formula \eqref{eq:q2qv} formulated in the collinear hadron frame. In this frame, the naive (unsubtracted and unrenormalized) definition is given by \cite{Bacchetta:2006tn,Mulders:1995dh,Goeke:2005hb}
\begin{eqnarray}
    \Delta^q_{ij}(z,\bm{p}_T) & = & \frac{1}{N_c}\sumint\int_{-\infty}^\infty\tfrac{\dd \lambda}{2\pi}\int \dd^{d-2}\bm{z}_T\,\e^{-i\frac{\lambda}{z}}\e^{i\bm{p}_T\cdot\bm{z}_T}\langle 0|\,\mathcal{W}_{\mathrm{SIDIS}}\,q_i(0)\,|P_h;X\rangle\langle P_h;X|\,\bar{q}_j(\lambda\,n+z_T)\,\mathcal{W}_{\mathrm{SIDIS}}\,|0\rangle\nn\\
    &=&z\,\left[\slash{P}_h\,D_1^q+\frac{i\,[\slash{p}_T,\slash{P}_h]}{2\,M_h}\,H_1^{\perp,q}+M_h\,E^q+\frac{x_B\,M_h}{z_h\,Q^2}\,i\,[\slash{P}_h,\slash{P}]\,H^q+...\right]\,.\label{eq:TMDDelta}
\end{eqnarray}
It is well-known that the Wilson line $\mathcal{W}_{\mathrm{SIDIS}}$ is non-trivial and process dependent due to the transverse quark momentum $\bm{p}_T$ in the correlator $\Delta$ in \eqref{eq:TMDDelta}. The exact form $\mathcal{W}_{\mathrm{SIDIS}}$ can be found in Refs.~\cite{Bacchetta:2006tn,Collins:2004nx}, but it is not relevant for this work. We further emphasize that the light-cone vector $n^\mu$ in \eqref{eq:TMDDelta} in the collinear hadron frame reads $n^\mu=\frac{2\,x_B}{z_h\,Q^2}P^\mu$, as discussed in \eqref{eq:LCvector}. The (naive) TMD FFs in the second line of \eqref{eq:TMDDelta} are the ordinary fragmentation function $D_1^q(z,-z^2\,\bm{p}_T^2)$ of an unpolarized quark fragmenting into an unpolarized hadron, the Collins fragmentation function $H_1^{\perp,q}(z,-z^2\,\bm{p}_T^2)$ \cite{Collins:1992kk}, and higher twist TMD fragmentation functions $E^q(z,-z^2\,\bm{p}_T^2)$ and $H^q(z,-z^2\,\bm{p}_T^2)$. We only presented the chiral-odd structures in the second line of \eqref{eq:TMDDelta} (apart from the $D_1^q$ structure) and neglected other structures that are irrelevant for this paper.

In the next step, we Lorentz-transform Eq.~\eqref{eq:TMDDelta} into the Breit frame. This concerns, in particular, the second line of \eqref{eq:TMDDelta}. As discussed in \eqref{eq:relpT}, we can identify $\bm{p}_{T,\mathrm{Br}}= (\bm{p}_T)_\perp$ for the $d-2$ dimensional transverse vectors, but for the $d$ dimensional version, we obtain
\begin{eqnarray}
     p_{T,\mathrm{Br}}^\mu = (p_T)_\perp^\mu - 2 x_B\,\tfrac{{(\bm{p}_T)^2_\perp}}{Q^2}\,P^\mu\,,\label{eq:relpT2}
\end{eqnarray}
where we replaced $\bm{\chi}_\perp=-(\bm{p}_T)_\perp/Q$ due to the delta function in \eqref{eq:q2qvBr}. Note that the second term in \eqref{eq:relpT2} is not to be neglected; it ensures that kinematical twist-3 contributions do appear in our final formulae. The form of the momentum $P_h^\mu$ of the detected hadron in the Breit frame has been discussed in \eqref{eq:PhBr}. By again using $\bm{\chi}_\perp=-(\bm{p}_T)_\perp/Q$, we obtain
\begin{equation}
    P_{h,\mathrm{Br}}^\mu = z_h\,\tfrac{(\bm{p}_T)_\perp^2}{Q^2}\,x_BP^\mu+z_h(x_BP^\mu+q^\mu)-z_h\,(p_T)_\perp^\mu\,.\label{eq:PhBrv}
\end{equation}
Eqs.~\eqref{eq:relpT2} and \eqref{eq:PhBrv}, inserted into the parameterization \eqref{eq:TMDDelta} with a subsequent integration over $(\bm{p}_T)_\perp$, simplify the input for the fragmentation process in Eq.~\eqref{eq:q2qvInt}
\begin{eqnarray}
    \left(\int \dd^{d-2}(\bm{p}_T)_\perp\,\Delta^q(z_h,(\bm{p}_T)_\perp)\right)_{\mathrm{Br}}&=& z_h^{2\epsilon}\left[(x_B\slash{P}+\slash{q})\,D_1^q(z_h)+M_h\,\frac{E^q(z_h)}{z_h}\right.\nn\\
    &&\quad\left.-2\,x_B\,M_h\,\frac{i[\slash{P},\slash{q}]}{Q^2}\,\left(\frac{H^q(z_h)}{2\,z_h}+(1-\epsilon)\,H_1^{\perp (1),q}(z_h)\right)+...\right]\,.\label{eq:intTMD}
\end{eqnarray}
In deriving the form \eqref{eq:intTMD}, we dropped terms linear in $(\bm{p}_T)_\perp$\footnote{The corresponding $(\bm{p}_T)_\perp$-integrals vanish since $\int \dd ^{d-2}(\bm{p}_T)_\perp\,(\bm{p}_T)_\perp\,D(z,-z^2(\bm{p}_T)_\perp^2) = 0$.}. Furthermore, the collinear intrinsic functions $D_1$, $E$, and $H$ are generically obtained through
\begin{equation}
\int\dd ^{d-2}(\bm{p}_T)_\perp\,D(z,-z^2\,(\bm{p}_T)_\perp^2)=z^{-2+2\epsilon}\,D(z)\,,\label{eq:intD}
\end{equation}
and the first TMD moment of the Collins function through
\begin{equation}
    \int \dd ^{d-2}(\bm{p}_T)_\perp\,(\bm{p}_T)_\perp^2\,H_1^{\perp q}(z,-z^2\,(\bm{p}_T)_\perp^2)=z^{-2+2\epsilon}\,2(1-\epsilon)\,M_h^2\,H_1^{\perp (1),q}(z)\,.\label{eq:H1perp1}
\end{equation}
Interestingly, the integrated correlator on the left hand side of \eqref{eq:intTMD}, transformed to the Breit frame, is described on the right hand side by both intrinsic, $E$, $H$, and kinematical, $H_1^{\perp(1)}$ twist-3 fragmentation functions. In particular, the functions $H$ and $H_1^{\perp(1)}$ enter into a combination such that an equation-of-motion relation can be directly applied. We will discuss this relation further below.\\

Let us test the factorization formula \eqref{eq:q2qvInt} for $n=0$ unobserved partons in the final state by calculating the LO contribution. At LO, the (amputated) hard scattering amplitude $\mathcal{M}^\mu_{q\to q,v,\mathrm{LO}}$ is simply given by the perturbative quark-photon vertex, i.e., $\mathcal{M}^\mu_{q\to q,v,\mathrm{LO}}=\gamma^\mu$ and $\overline{\mathcal{M}}^\nu_{q\to q,v,\mathrm{LO}}=\gamma^\nu$. By combining these LO amplitudes with \eqref{eq:q2qvInt}, \eqref{eq:intTMD}, and \eqref{eq:Phi}, we identify three relevant covariant LO expressions for the hadronic tensor;
\begin{eqnarray}
    \tilde{W}^{\mu\nu}_{\mathrm{LO,\,unpol}} & = & \dd z_h\,\left[-g_\perp^{\mu\nu}\right]\,\sum_q\,e_q^2\,f_1^q(x_B)\,D_1^q(z_h)\,,\label{eq:WLOunpol}\\
    \tilde{W}^{\mu\nu}_{\mathrm{LO,q\to q,\,sym}}(S) & = & \dd z_h\,\left[\tfrac{8\,M_h\,x_B^2}{Q^4}\,P^{\{\mu}\,\epsilon^{\nu\}PqS}\right]\sum_q e_q^2\,h_1^q(x_B)\,\left(\tfrac{H^q(z_h)}{2\,z_h}+(1-\epsilon)\,H_1^{\perp (1),q}(z_h)\right) \,,\label{eq:WLOkinsym}\\
        \tilde{W}^{\mu\nu}_{\mathrm{LO,q\to q,\,anti}}(S) & = & \dd z_h\,\left[\tfrac{-8\,i\,M_h\,x_B^2}{Q^4}\,P^{[\mu}\,\epsilon^{\nu ]PqS}\right]\sum_q e_q^2\,h_1^q(x_B)\,\left(\tfrac{E^q(z_h)}{2\,z_h}\right) \,.\label{eq:WLOkinanti}
\end{eqnarray}
The tensor $\tilde{W}^{\mu\nu}_{\mathrm{LO,\,unpol}}$ constitutes the full LO result for spin-independent, $\bm{P}_{h\perp}$-integrated SIDIS. It is symmetric under an exchange of indices $\mu\leftrightarrow\nu$. Note that it satisfies electromagnetic current conservation $q_\mu \tilde{W}^{\mu\nu}_{\mathrm{LO,\,unpol}}=q_{\nu}\tilde{W}^{\mu\nu}_{\mathrm{LO,\,unpol}}=0$. The nucleon spin-dependent symmetric and anti-symmetric hadronic tensors are given in \eqref{eq:WLOkinsym} and \eqref{eq:WLOkinanti}, respectively\footnote{The notation in \eqref{eq:WLOkinsym}, $a^{\{\mu} b^{\nu\}}\equiv a^\mu b^\nu+a^\nu b^\mu$, indicates symmetrization, and $a^{[\mu} b^{\nu]}\equiv a^\mu b^\nu-a^\nu b^\mu$ in \eqref{eq:WLOkinanti} indicates anti-symmetrization.}. They do not satisfy e.m. current conservation; therefore, they cannot be considered the complete result. Contributions from dynamical quark-gluon-quark fragmentation are missing at this point. We will discuss those contributions below, but before we do so, we will first focus on contributions with $n=1$ unobserved partons in the final state.
\subsubsection{Real contributions, $n=1$}
The factorization setup in the collinear hadron frame is provided in \eqref{eq:q2qr} and is sketched in the right panel of Fig.~\ref{fig:2Parton}. The delta function in \eqref{eq:q2qr} can be expressed as
\begin{eqnarray}
    \delta((\bar{k}+q-\bar{p})^2) &= &\tfrac{1}{v}\,\delta(\tfrac{1-v}{v}\tfrac{1-w}{w}Q^2-(\bm{q}_T-\tfrac{1}{v}\bm{p}_T)^2)\nn\\
     & \to & \tfrac{z_h^2}{v}\,\delta(\tfrac{1-v}{v}\tfrac{1-w}{w}\,z_h^2Q^2-(\bm{P}_{h\perp}+\tfrac{z_h}{v}(\bm{p}_T)_\perp)^2)\,,\label{eq:deltar}
\end{eqnarray}
where in the second line, we transformed the delta function into the Breit frame by means of Eqs.~\eqref{eq:relqTPhp}, \eqref{eq:relpT}. The approximated parton momenta $\bar{k}^\mu$ in \eqref{eq:kcollapp} and $\bar{p}^\mu$ in \eqref{eq:pcollapp} take the following form in the Breit frame (upon implementing the constraint enforced by the delta function \eqref{eq:deltar}),
\begin{eqnarray}
    \bar{k}^\mu & = & \tfrac{x_B}{w}P^\mu_{\mathrm{Br}}\,,\nn\\
    \bar{p}^\mu & = & v\,\left[1+\tfrac{1-v}{v}\tfrac{1-w}{w}\right]\,x_B\,P^\mu_{\mathrm{Br}}+v\,q_{\mathrm{Br}}^\mu+\tfrac{v}{z_h}(P_{h\perp}+\tfrac{z_h}{v}(p_T)_\perp)^\mu\,,\nn\\
    P_h^\mu & = & z_h\,\bm{\chi}_\perp^2\,x_B\,P^\mu + z_h\,(x_B\,P^\mu+q^\mu)+P_{h\perp}^\mu\,.\label{eq:kbPhBr}
\end{eqnarray}
Then, we consider the $\bm{P}_{h\perp}$-integrated hadronic tensor \eqref{eq:q2qr} in the Breit frame, analogous to \eqref{eq:q2qvInt},
\begin{eqnarray}
    \tilde{W}^{\mu\nu}_{q\to q,r}&\equiv& \frac{\dd z_h}{2\,z_h}\int \dd^{d-2}\bm{P}_{h\perp}\, W^{\mu\nu}_{q\to q,r,\mathrm{Br}} \nn\\
    &=&\dd z_h\,\int_{x_B}^1 \tfrac{\dd w}{w}\int_{z_h}^1\tfrac{\dd v}{v}\,\tfrac{x_B}{2\,w}\int \dd^{d-2}(\bm{p}_T)_\perp\,\int \dd^{d-2}\bm{P}_{h\perp}\delta\left(\tfrac{1-v}{v}\tfrac{1-w}{w}\,z_h^2Q^2-(\bm{P}_{h\perp}+\tfrac{z_h}{v}(\bm{p}_T)_\perp)^2\right)\times\nn\\
    &&\tfrac{1}{(2\pi)^{d-1}}\sum_qe_q^2\,\sum_{I_1}\mathrm{Tr}\left[\overline{\mathcal{M}}_{q\to q,r}^\nu\,(\bar{k},\bar{p},q)\,\Delta^q(\tfrac{z_h}{v},(\bm{p}_T)_\perp)\,\mathcal{M}^\mu_{q\to q,r}(\bar{k},\bar{p},q)\,\Phi^q(\tfrac{x_B}{w})\right]\,.\label{eq:q2qr2}
\end{eqnarray}
The form of the $\bm{P}_{h\perp}$-integrated hadronic tensor \eqref{eq:q2qr2} allows for a simple treatment of the transverse momentum integrals. In particular, we can perform a shift 
\begin{equation}
    \bm{P}_{h\perp}\to \tilde{\bm{P}}_{h\perp}\equiv \bm{P}_{h\perp}+\tfrac{z_h}{v}(\bm{p}_T)_\perp\,,\label{eq:shift}
\end{equation}
and obtain the final form of the factorization formula for real corrections,
\begin{eqnarray}
    \tilde{W}^{\mu\nu}_{q\to q,r}
    &=&\dd z_h\,\int_{x_B}^1 \tfrac{\dd w}{w}\int_{z_h}^1\tfrac{\dd v}{v}\,\tfrac{x_B}{2\,w}\int \dd^{d-2}\tilde{\bm{P}}_{h\perp}\,\delta\left(\tfrac{1-v}{v}\tfrac{1-w}{w}\,z_h^2Q^2-\tilde{\bm{P}}_{h\perp}^2\right)\times\nn\\
    &&\sum_q\tfrac{e_q^2}{(2\pi)^{d-1}}\,\sum_{I_1}\mathrm{Tr}\left[\overline{\mathcal{M}}_{q\to q,r}^\nu\,(\bar{k},\tilde{p},q)\,\left(\int \dd^{d-2}(\bm{p}_T)_\perp\,\Delta^q(\tfrac{z_h}{v},(\bm{p}_T)_\perp)\right)\,\mathcal{M}^\mu_{q\to q,r}(\bar{k},\tilde{p},q)\,\Phi^q(\tfrac{x_B}{w})\right]\,,\label{eq:q2qrInt}
\end{eqnarray}
with shifted new four momenta
\begin{eqnarray}
    \tilde{p}^\mu & = & v\,\left[1+\tfrac{1-v}{v}\tfrac{1-w}{w}\right]\,x_B\,P^\mu_{\mathrm{Br}}+v\,q_{\mathrm{Br}}^\mu+\tfrac{v}{z_h}\tilde{P}_{h\perp}^\mu\,,\nn\\
    \tilde{P}_h^\mu & = & z_h\,\tfrac{(\tilde{\bm{P}}_{h\perp}-\tfrac{z_h}{v}(\bm{p}_T)_\perp)^2}{z_h^2\,Q^2}\,x_B\,P^\mu + z_h\,(x_B\,P^\mu+q^\mu)+\tilde{P}_{h\perp}^\mu-\tfrac{z_h}{v}(p_T)_\perp^\mu\,.\label{eq:pPhshift}
\end{eqnarray}
The shift \eqref{eq:shift} guaranties that the momentum $\tilde{p}^\mu$ of the fragmenting quark in \eqref{eq:pPhshift} does not depend on the transverse momentum $(\bm{p}_T)_\perp$. Consequently, the partonic amplitude $\mathcal{M}^\mu_{q\to q,r}$ does not depend on $(\bm{p}_T)_\perp$ either, and we can directly implement the $(\bm{p}_T)_\perp$-integrated correlator $\int \dd^{d-2}(\bm{p}_T)_\perp\,\Delta^q(\tfrac{z_h}{v},(\bm{p}_T)_\perp)$ into the formula \eqref{eq:q2qrInt}. As for the virtual corrections \eqref{eq:intTMD}, we have to be careful with the parameterization. By means of the momentum $\tilde{P}_{h\perp}^\mu$ in \eqref{eq:pPhshift}, the Breit frame expression \eqref{eq:relpT2}, and the fact that terms linear in $(\bm{p}_T)_\perp$ drop out, we can show that the integrated correlator in \eqref{eq:q2qrInt} simplifies effectively as follows,
\begin{eqnarray}
    \left(\int \dd^{d-2}(\bm{p}_T)_\perp\,\Delta^q(\tfrac{z_h}{v},(\bm{p}_T)_\perp)\right)&\to& z_h^{2\epsilon}v^{-2\epsilon}\left[\tfrac{z_h}{v}\tilde{\slash{p}}\,D_1^q(\tfrac{z_h}{v})+M_h\,\frac{E^q(\tfrac{z_h}{v})}{\tfrac{z_h}{v}}\right.\nn\\
    &&\quad\left.-2\,\tfrac{w}{v}\,M_h\,\frac{i[\bar{\slash{k}},\tilde{\slash{p}}]}{Q^2}\,\left(\frac{H^q(\tfrac{z_h}{v})}{2\,\tfrac{z_h}{v}}+(1-\epsilon)\,H_1^{\perp (1),q}(\tfrac{z_h}{v})\right)+...\right]\,.\label{eq:intTMDr}
\end{eqnarray}

In the same way as we did for the LO expressions \eqref{eq:WLOunpol}, \eqref{eq:WLOkinsym}, \eqref{eq:WLOkinanti}, we can identify three hadronic tensors that are relevant to this paper,
\begin{eqnarray}
    \tilde{W}^{\mu\nu}_{\mathrm{NLO,q\to q,real,unpol}}&=& \dd z_h\,\int_{x_B}^1 \tfrac{\dd w}{w}\int_{z_h}^1\tfrac{\dd v}{v}\,\sum_q e_q^2\,\tfrac{z_h}{4\,v}\,f_1^q(\tfrac{x_B}{w})\,D_1^q(\tfrac{z_h}{v})\,z_h^{2\epsilon}v^{-2\epsilon}\times\label{eq:Wrealunpol}\\
    &&\hspace{-2cm}\int \tfrac{\dd^{d-2}\tilde{\bm{P}}_{h\perp}}{(2\pi)^{d-1}}\,\delta\left(\tfrac{1-v}{v}\tfrac{1-w}{w}\,z_h^2 Q^2-\tilde{\bm{P}}_{h\perp}^2\right)\,\sum_{I_1}\mathrm{Tr}\left[\overline{\mathcal{M}}_{q\to q,r}^\nu\,(\bar{k},\tilde{p},q)\,\tilde{\slash{p}}\,\mathcal{M}^\mu_{q\to q,r}(\bar{k},\tilde{p},q)\,\slash{k}\right]\,,\nn\\
    \tilde{W}^{\mu\nu}_{\mathrm{NLO,q\to q,real,sym}}(S)&=& \dd z_h\,\int_{x_B}^1 \tfrac{\dd w}{w}\int_{z_h}^1\tfrac{\dd v}{v}\,\sum_q e_q^2\,(\tfrac{w\,M_h}{4\,v\,Q^2})\,h_1^q(\tfrac{x_B}{w})\,\left(\frac{H^q(\tfrac{z_h}{v})}{2\,\tfrac{z_h}{v}}+(1-\epsilon)\,H_1^{\perp (1),q}(\tfrac{z_h}{v})\right) \,z_h^{2\epsilon}v^{-2\epsilon}\times\label{eq:Wrealsym}\\
    &&\hspace{-2cm}\int\tfrac{\dd^{d-2}\tilde{\bm{P}}_{h\perp}}{(2\pi)^{d-1}}\,\delta\left(\tfrac{1-v}{v}\tfrac{1-w}{w}\,z_h^2Q^2-\tilde{\bm{P}}_{h\perp}^2\right)\,\sum_{I_1}\mathrm{Tr}\left[\overline{\mathcal{M}}_{q\to q,r}^\nu\,(\bar{k},\tilde{p},q)\,[\bar{\slash{k}},\tilde{\slash{p}}]\,\mathcal{M}^\mu_{q\to q,r}(\bar{k},\tilde{p},q)\,[\bar{\slash{k}},\slash{S}](i\gamma_5)\right]\,,\nn\\
    \tilde{W}^{\mu\nu}_{\mathrm{NLO,q\to q,real,anti}}(S)&=& \dd z_h\,\int_{x_B}^1 \tfrac{\dd w}{w}\int_{z_h}^1\tfrac{\dd v}{v}\,\sum_q e_q^2\,(\tfrac{i\,M_h}{8})\,h_1^q(\tfrac{x_B}{w})\,\frac{E^q(\tfrac{z_h}{v})}{\tfrac{z_h}{v}} \,z_h^{2\epsilon}v^{-2\epsilon}\times\label{eq:Wrealanti}\\
    &&\hspace{-2cm}\int\tfrac{\dd^{d-2}\tilde{\bm{P}}_{h\perp}}{(2\pi)^{d-1}}\,\delta\left(\tfrac{1-v}{v}\tfrac{1-w}{w}\,z_h^2Q^2-\tilde{\bm{P}}_{h\perp}^2\right)\,\sum_{I_1}\mathrm{Tr}\left[\overline{\mathcal{M}}_{q\to q,r}^\nu\,(\bar{k},\tilde{p},q)\,\mathcal{M}^\mu_{q\to q,r}(\bar{k},\tilde{p},q)\,[\bar{\slash{k}},\slash{S}](i\gamma_5)\right]\,.\nn
\end{eqnarray}

Interestingly, the last remaining integration over the transverse hadron momentum $\tilde{\bm{P}}_{h\perp}$ in Eqs.~\eqref{eq:Wrealunpol}, \eqref{eq:Wrealsym}, \eqref{eq:Wrealanti} turns out to be simple since the structure of the momentum $\tilde{p}$ in \eqref{eq:pPhshift} prevents any scalar products $\tilde{p}^2(=0)$, $\bar{k}\cdot\tilde{p}$, $q\cdot\tilde{p}$ from depending on the integrated momentum $\tilde{\bm{P}}_{h\perp}$. As a result, only two simple integrals need to be evaluated (with $S_\epsilon=(4\pi)^\epsilon/\Gamma(1-\epsilon)$),
\begin{eqnarray}
   \mu^{2\epsilon} \int\tfrac{\dd^{d-2}\tilde{\bm{P}}_{h\perp}}{(2\pi)^{d-1}}\,\delta\left(\tfrac{1-v}{v}\tfrac{1-w}{w}\,z_h^2Q^2-\tilde{\bm{P}}_{h\perp}^2\right)\,  & = & \tfrac{S_\epsilon}{8\pi^2} (1-v)^{-\epsilon}\,(1-w)^{-\epsilon}\,v^\epsilon\,w^\epsilon\,z_h^{-2\epsilon}\,\left(\tfrac{Q^2}{\mu^2}\right)^{-\epsilon},\label{eq:PhpInts}\\
    \mu^{2\epsilon} \int\tfrac{\dd^{d-2}\tilde{\bm{P}}_{h\perp}}{(2\pi)^{d-1}}\,\delta\left(\tfrac{1-v}{v}\tfrac{1-w}{w}\,z_h^2Q^2-\tilde{\bm{P}}_{h\perp}^2\right)\, \tilde{\bm{P}}_{h\perp}^\alpha \tilde{\bm{P}}_{h\perp}^\beta & = & -\tfrac{z_h^2\,Q^2}{2(1-\epsilon)}\,g_{\perp}^{\alpha\beta}\,\tfrac{S_\epsilon}{8\pi^2} (1-v)^{1-\epsilon}\,(1-w)^{1-\epsilon}\,v^{-1+\epsilon}\,w^{-1+\epsilon}\,z_h^{-2\epsilon}\,\left(\tfrac{Q^2}{\mu^2}\right)^{-\epsilon}.\nn
\end{eqnarray}
The powers $(1-v)^{-\epsilon}$, $(1-w)^{-\epsilon}$ that the $d-2$-dimensional $\tilde{\bm{P}}_{h\perp}$ integration provides are important for regularizing soft and collinear divergences that are well-known to appear in inclusive partonic cross sections. To be specific, the following expansions in $\epsilon$ need to be applied eventually,
\begin{eqnarray}
    \frac{1}{(1-w)^{1+\epsilon}} & = & -\frac{1}{\epsilon}\delta(1-w)+\frac{1}{(1-w)_+}-\epsilon\,\left(\frac{\ln(1-w)}{1-w}\right)_++\mathcal{O}(\epsilon^2)\,,\nn\\
     \frac{1}{(1-v)^{1+\epsilon}} & = & -\frac{1}{\epsilon}\delta(1-v)+\frac{1}{(1-v)_+}-\epsilon\,\left(\frac{\ln(1-v)}{1-v}\right)_++\mathcal{O}(\epsilon^2)\,,\nn\\
     \frac{1}{(1-v)^{1+\epsilon}(1-w)^{1+\epsilon}} & = & \frac{1}{\epsilon^2}\delta(1-v)\,\delta(1-w)-\frac{1}{\epsilon}\frac{\delta(1-v)}{(1-w)_+}-\frac{1}{\epsilon}\frac{\delta(1-w)}{(1-v)_+}\nn\\
     &&+\left(\frac{\ln(1-w)}{1-w}\right)_+\delta(1-v)+\left(\frac{\ln(1-v)}{1-v}\right)_+\delta(1-w)+\frac{1}{(1-v)_+\,(1-w)_+}+\mathcal{O}(\epsilon)\,.\label{eq:epsExp}
\end{eqnarray}
Those expressions \eqref{eq:epsExp} contain the usual plus-prescriptions (see, e.g., also Ref.~\cite{deFlorian:1997zj}), 
\begin{eqnarray}
    \int_{x_B}^1\dd w\,\frac{f(w)}{(1-w)_+}&=&\int_{x_B}^1\dd w\,\frac{f(w)-f(1)}{1-w}+f(1)\,\ln(1-x_B)\,,\nn\\
     \int_{x_B}^1\dd w\,\left(\frac{\ln(1-w)}{1-w}\right)_+f(w) & = & \int_{x_B}^1\dd w\,\frac{\ln(1-w)}{1-w}(f(w)-f(1))+\tfrac{1}{2}f(1)\,\ln^2(1-x_B)\,,\nn\\ 
     \int_{z_h}^1\dd v\int_{x_B}^1\dd w\,\frac{f(v,w)}{(1-v)_+\,(1-w)_+} & = & \int_{z_h}^1\dd v\int_{x_B}^1\dd w\,\frac{[f(v,w)-f(1,w)-f(v,1)+f(1,1)]}{(1-v)\,(1-w)}\nn\\
     &&+\ln(1-x_B)\int_{z_h}^1 \dd v \frac{f(v,1)-f(1,1)}{1-v}+\ln(1-z_h)\int_{x_B}^1 \dd w \frac{f(1,w)-f(1,1)}{1-w}\nn\\
     &&+f(1,1)\,\ln(1-x_B)\,\ln(1-z_h)\,.\label{eq:plusprescriptions}
\end{eqnarray}

The first two lines are given for the plus-prescriptions for the variable $w$, but they may also be used for the variable $v$ (and $z_h$ instead of $x_B$).

\subsection{3-Parton correlations}
\begin{figure}
    \centering
    \includegraphics[width=0.47\linewidth]{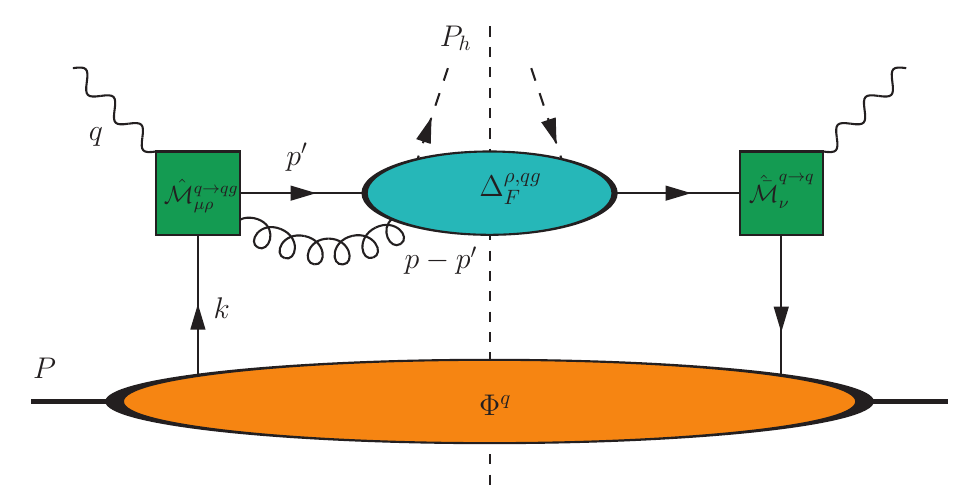}\,\,
     \includegraphics[width=0.47\linewidth]{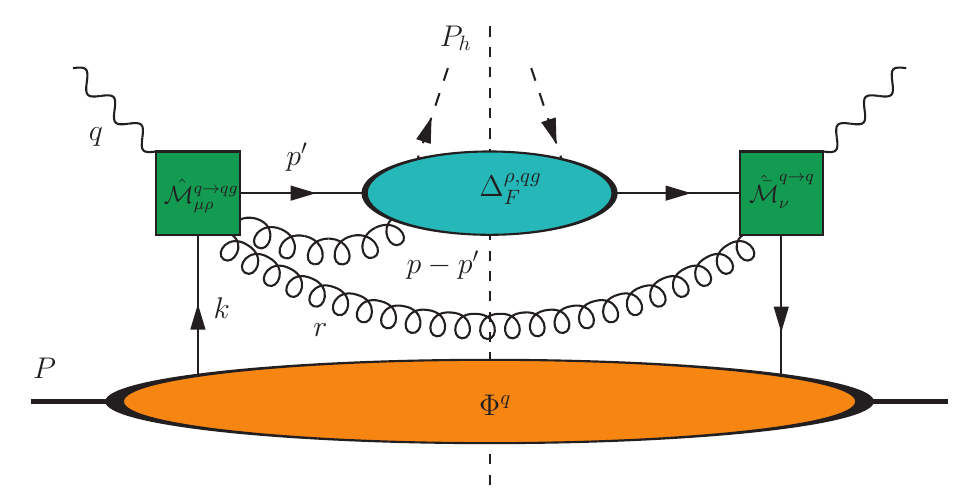}
    \caption{Collinear twist-3 factorization setup: dynamical quark-gluon-quark fragmentation, \textbf{left:} $n=0$ unobserved partons in the final state, \textbf{right:} $n=1$ unobserved gluon in the final state.}
    \label{fig:3Parton}
\end{figure}
The \textit{dynamical} twist-3 fragmentation contributions may be handled in the same way as we did for the \textit{intrinsic} and \textit{kinematic} twist-3 contributions in the previous section. Up to NLO accuracy, only two setups of the collinear twist-3 factorization formulae need to be considered: one without an unobserved parton in the final state (\textit{virtual} contributions) and one with an unobserved parton (\textit{real} contributions). Both setups are depicted in Fig.~\ref{fig:3Parton}. In essence, one follows the same steps as described in the previous section: first, we formulate the factorization in the collinear hadron frame \eqref{eq:collframe} assuming a light-cone gauge \eqref{eq:LCgauge}, and then we perform the collinear approximation for the parton momenta in the hard part as in \eqref{eq:kcollapp}, \eqref{eq:pcollapp}. We may drop all transverse momentum dependencies right away and set $p_T=0$. However, since we encounter an additional parton momentum $p^\prime$ in the fragmentation process, we apply an additional collinear approximation,
\begin{equation}
    p^{\prime \mu} \simeq \tfrac{1}{z^\prime}P_h^\mu\equiv \tfrac{\zeta\,v}{z_h}\,P_h^\mu=\zeta\,\bar{p}^\mu\Big|_{p_T=0}\,.\label{eq:ppcollapp}
\end{equation}
Consequently, the factorization formula will contain an extra integration over the parameter $\zeta$. The notation has been adopted from Ref.~\cite{Gamberg:2018fwy} (with $\zeta$ called $\beta$ in this reference), and the support properties of quark-gluon-quark fragmentation functions demand $0\le \zeta\le 1$.

Then, analogously to the previous section, we perform a Lorentz-transform to the Breit frame \eqref{eq:Brframe} and integrate out the $d-2$-dimensional transverse hadron momentum $\bm{P}_{h\perp}$. In the following, we provide the factorization formula as in \eqref{eq:q2qvInt} and \eqref{eq:q2qrInt} for the dynamical fragmentation contributions without going into detail,
\begin{eqnarray}
    \tilde{W}^{\mu\nu}_{q\to qg,v} & = & \dd z_h\,z_h^{-2\epsilon}\tfrac{x_B}{Q^2}\,\int_{0}^1\dd \zeta\,\frac{(-i)g_{\perp,\rho\sigma}}{1-\zeta}\times\nn\\
    &&\sum_q e_q^2\,\mathrm{Tr}\left[\overline{\mathcal{M}}^{\nu}_{q\to q,v}(\bar{k}_{\mathrm{Br}},\bar{p}_{\mathrm{Br}},\bar{q}_{\mathrm{Br}})\,\Delta_F^{q;\rho}(z_h,\zeta)\,\mathcal{M}^{\mu;\sigma}_{q\to qg,v}(\bar{k}_{\mathrm{Br}},\bar{p}_{\mathrm{Br}},\zeta\,\bar{p}_{\mathrm{Br}},\bar{q}_{\mathrm{Br}})\,\Phi^q(x_B)\right]+\mathrm{c.c.}\,,\label{eq:Wq2qgvInt}
\end{eqnarray}
for the virtual contributions (see Fig.~\ref{fig:3Parton}, left panel), and
\begin{eqnarray}
    \tilde{W}^{\mu\nu}_{q\to qg,r}
    &=&\dd z_h\,\int_{x_B}^1 \tfrac{\dd w}{w}\int_{z_h}^1\tfrac{\dd v}{v}\,\int_0^1 \dd \zeta\,\frac{(-i)g_{\perp,\rho\sigma}}{1-\zeta}\,\frac{x_B}{2\,w}\,\int \tfrac{\dd^{d-2}\bm{P}_{h\perp}}{(2\pi)^{d-1}}\,\delta\left(\tfrac{1-v}{v}\tfrac{1-w}{w}\,z_h^2Q^2-\bm{P}_{h\perp}^2\right)\times\nn\\
    &&\sum_qe_q^2\,\sum_{I_1}\mathrm{Tr}\left[\overline{\mathcal{M}}_{q\to q,r}^\nu\,(\bar{k},\bar{p},q)\,\Delta_F^{q;\rho}(\tfrac{z_h}{v},\zeta)\,\mathcal{M}^{\mu;\sigma}_{q\to qg,r}(\bar{k},\bar{p},\zeta\,\bar{p},q)\,\Phi^q(\tfrac{x_B}{w})\right]+\mathrm{c.c.}\,,\label{eq:Wq2qgrInt}
\end{eqnarray}
for the real contributions (see Fig.~\ref{fig:3Parton}, right panel).
The momenta $\bar{k}_{\mathrm{Br}}$, $\bar{p}_{\mathrm{Br}}$, $\bar{q}_{\mathrm{Br}}$ that appear in the hard scattering amplitudes of the virtual contribution \eqref{eq:Wq2qgvInt} are given in Eqs.~\eqref{eq:Brframe} and \eqref{eq:pBr}. On the other hand, the momenta $\bar{k},\,\bar{p}$ in the hard scattering amplitudes of the real contribution \eqref{eq:Wq2qgrInt} are provided by \eqref{eq:kcollapp}, \eqref{eq:pcollapp} and \eqref{eq:ppcollapp}. Again, we note that all dependencies on transverse parton momenta are dropped; hence, a shift like \eqref{eq:pPhshift} is not necessary.

The key ingredient of the dynamical contributions \eqref{eq:Wq2qgvInt}, \eqref{eq:Wq2qgrInt} is the \textit{quark-gluon-quark}($qgq$) fragmentation correlator $\Delta_F^{q;\rho}(z,\zeta)$. This object has been discussed at length in Refs.~\cite{Gamberg:2018fwy} (see also \cite{Kanazawa:2015ajw}), and we will only provide the key points that are necessary for this paper. The definition and parameterization of the $qgq$ fragmentation correlator read in $d=4-2\epsilon$ dimensions,
\begin{eqnarray}
\Delta^{q;\rho}_{F;ij}(z,\zeta) & = &   \frac{1}{N_c}\sumint\int_{-\infty}^\infty\tfrac{\dd \lambda}{2\pi}\int_{-\infty}^\infty\tfrac{\dd \mu}{2\pi}\,\e^{-i\tfrac{\lambda}{z}\zeta}\e^{-i\tfrac{\mu}{z}(1-\zeta)}\langle0|\,[\infty\, n; 0]\,q_i(0)\,|P_h X\rangle\times\nn\\
&&\hspace{3cm}\langle P_hX|\,\bar{q}_j(\lambda\,n)\,[\lambda\,n;\mu\,n]\,ig\mu_R^\epsilon\,F^{n\rho}(\mu\,n)\,[\mu\,n;\infty\,n]\,|0\rangle   \nn\\
&= & M_h\,z^{-1+2\epsilon}\left[\tfrac{i}{2}\left([\slash{P}_h,\gamma^\rho]-\tfrac{2x_B}{z_h\,Q^2}[\slash{P}_h,\slash{P}]\,P_h^\rho \right)\,\left(\mathrm{Im}\hat{H}^{qg}_{FU}(z,\zeta)+i\,\mathrm{Re}\hat{H}^{qg}_{FU}(z,\zeta)\right)+...\right]\,.\label{eq:Deltaqgq}
\end{eqnarray}
The field-theoretical definition of the object $\Delta_F$ in the first line of \eqref{eq:Deltaqgq} contains not only the quark fields $q$ and $\bar{q}$ but also the gluonic field-strength tensor $F^{n\rho}\equiv n_\kappa F^{\kappa\rho}$, accompanied by the strong coupling constant $g$ and the renormalization scale $\mu_R$. Note that a factor $i\mu_R^\epsilon g\,t^\alpha$ ($t^\alpha$ being a color matrix) in the definition \eqref{eq:Deltaqgq} typically originates from the hard scattering amplitudes. In other words, the partonic amplitudes $\mathcal{M}^{\mu;\sigma}_{q\to qg,v}$ and $\mathcal{M}^{\mu;\sigma}_{q\to qg,r}$ in \eqref{eq:Wq2qgvInt} and \eqref{eq:Wq2qgrInt} are to be understood as the (amputated) sums of Feynman diagrams with a factor $i\mu_R^\epsilon g\,t^\alpha$ separated out.

The parameterization in the last line of \eqref{eq:Deltaqgq} provides one $qgq$ fragmentation function $\hat{H}^{qg}_{FU}$ for an unpolarized detected hadron\footnote{Other spin-dependent fragmentation functions are denoted by $...$ . They are irrelevant for this work.}. Unlike $qgq$ correlation functions in the nucleon, the $qgq$ fragmentation functions in \eqref{eq:Deltaqgq} are complex valued functions. The origin of this feature is that time-reversal cannot be applied to a fragmentation matrix element like \eqref{eq:Deltaqgq}, and hence does not constrain the $qgq$ fragmentation functions to be real. Another important difference compared to $qgq$ correlations in the nucleon is that so-called \textit{pole contributions} do not appear in the twist-3 fragmentation process \cite{Meissner:2008yf,Gamberg:2018fwy}. This means, in particular, that the function $\hat{H}_{FU}(z,\zeta)$ in \eqref{eq:Deltaqgq} vanishes at the boundaries $\zeta=0$ and $\zeta=1$, along with its derivatives,
\begin{equation}
    \hat{H}_{FU}(z,1)=\hat{H}_{FU}(z,0)=0\quad;\quad (\partial_\zeta\hat{H}_{FU})(z,1)=(\partial_\zeta\hat{H}_{FU})(z,0)=0\,.\label{eq:noPoles}
\end{equation}

The parameterization \eqref{eq:Deltaqgq}, along with \eqref{eq:Phi}, allows for a separation of the hadronic tensors \eqref{eq:Wq2qgvInt} and \eqref{eq:Wq2qgrInt} into symmetric and antisymmetric parts. For the virtual contributions, we obtain
\begin{eqnarray}
    \tilde{W}^{\mu\nu}_{\mathrm{NLO},q\to qg,v,\mathrm{sym}} & = & \dd z_h\,\left(\frac{-M_h}{16\,Q^2}\right)\,\int_{0}^1\dd \zeta\,\,\left(\sum_q e_q^2\,h_1^q(x_B)\,\frac{\mathrm{Im}\hat{H}^{qg}_{FU}(z_h,\zeta)}{1-\zeta}\right)\,g_{\perp,\rho\sigma}\times\nn\\
    &&\hspace{-3cm}\,\mathrm{Tr}\left[\overline{\mathcal{M}}^{\{\nu}_{q\to q,v}(\bar{k}_{\mathrm{Br}},\bar{p}_{\mathrm{Br}},\bar{q}_{\mathrm{Br}})\,\left([\slash{\bar{p}}_{\mathrm{Br}},\gamma^\rho]-\tfrac{2}{Q^2}[\slash{\bar{p}}_{\mathrm{Br}},\slash{\bar{k}}_{\mathrm{Br}}]\,\bar{p}_{\mathrm{Br}}^\rho\right)\,\mathcal{M}^{\mu\};\sigma}_{q\to qg,v}(\bar{k}_{\mathrm{Br}},\bar{p}_{\mathrm{Br}},\zeta\,\bar{p}_{\mathrm{Br}},\bar{q}_{\mathrm{Br}})\,[\slash{\bar{k}}_{\mathrm{Br}},\slash{S}]\gamma_5\right]+\mathrm{c.c.}\,,\label{eq:Wsymq2qgv}\\
     \tilde{W}^{\mu\nu}_{\mathrm{NLO},q\to qg,v,\mathrm{anti}} & = & \dd z_h\,\left(\frac{i\,M_h}{16\,Q^2}\right)\,\int_{0}^1\dd \zeta\,\,\left(\sum_q e_q^2\,h_1^q(x_B)\,\frac{\mathrm{Re}\hat{H}^{qg}_{FU}(z_h,\zeta)}{1-\zeta}\right)\,g_{\perp,\rho\sigma}\times\nn\\
    &&\hspace{-3cm}\,\mathrm{Tr}\left[\overline{\mathcal{M}}^{[\nu}_{q\to q,v}(\bar{k}_{\mathrm{Br}},\bar{p}_{\mathrm{Br}},\bar{q}_{\mathrm{Br}})\,\left([\slash{\bar{p}}_{\mathrm{Br}},\gamma^\rho]-\tfrac{2}{Q^2}[\slash{\bar{p}}_{\mathrm{Br}},\slash{\bar{k}}_{\mathrm{Br}}]\,\bar{p}_{\mathrm{Br}}^\rho\right)\,\mathcal{M}^{\mu];\sigma}_{q\to qg,v}(\bar{k}_{\mathrm{Br}},\bar{p}_{\mathrm{Br}},\zeta\,\bar{p}_{\mathrm{Br}},\bar{q}_{\mathrm{Br}})\,[\slash{\bar{k}}_{\mathrm{Br}},\slash{S}]\gamma_5\right]+\mathrm{c.c.}\,,\label{eq:Wantiq2qgv}
\end{eqnarray}
and for the real contributions,
\begin{eqnarray}
    \tilde{W}^{\mu\nu}_{\mathrm{NLO},q\to qg,r,\mathrm{sym}} & = & \dd z_h\,\left(\frac{-M_h}{32}\right)\int_{x_B}^1 \tfrac{\dd w}{w}\int_{z_h}^1\tfrac{\dd v}{v}\,\int_0^1 \dd \zeta\,\left(\sum_qe_q^2\,h_1^q(\tfrac{x_B}{w})\,\frac{\mathrm{Im}\hat{H}^{qg}_{FU}(\tfrac{z_h}{v},\zeta)}{1-\zeta}\right)\times\nn\\
    &&\int \tfrac{\dd^{d-2}\bm{P}_{h\perp}}{(2\pi)^{d-1}}\,\delta\left(\tfrac{1-v}{v}\tfrac{1-w}{w}\,z_h^2Q^2-\bm{P}_{h\perp}^2\right)\,g_{\perp,\rho\sigma}\,v^{-2\epsilon}z_h^{2\epsilon}\times\nn\\
    &&\sum_{I_1}\mathrm{Tr}\left[\overline{\mathcal{M}}_{q\to q,r}^{\{\nu}\,(\bar{k},\bar{p},q)\,\left([\slash{\bar{p}},\gamma^\rho]-\tfrac{2\,w}{v\,Q^2}[\slash{\bar{p}},\slash{\bar{k}}]\,\bar{p}^\rho\right)\,\mathcal{M}^{\mu\};\sigma}_{q\to qg,r}(\bar{k},\bar{p},\zeta\,\bar{p},q)\,[\slash{\bar{k}},\slash{S}]\gamma_5]\right]+\mathrm{c.c.}\,,\label{eq:Wsymq2qgr}\\
    \tilde{W}^{\mu\nu}_{\mathrm{NLO},q\to qg,r,\mathrm{anti}} & = & \dd z_h\,\left(\frac{i\,M_h}{32}\right)\int_{x_B}^1 \tfrac{\dd w}{w}\int_{z_h}^1\tfrac{\dd v}{v}\,\int_0^1 \dd \zeta\,\left(\sum_qe_q^2\,h_1^q(\tfrac{x_B}{w})\,\frac{\mathrm{Re}\hat{H}^{qg}_{FU}(\tfrac{z_h}{v},\zeta)}{1-\zeta}\right)\times\nn\\
    &&\int \tfrac{\dd^{d-2}\bm{P}_{h\perp}}{(2\pi)^{d-1}}\,\delta\left(\tfrac{1-v}{v}\tfrac{1-w}{w}\,z_h^2Q^2-\bm{P}_{h\perp}^2\right)\,g_{\perp,\rho\sigma}\,v^{-2\epsilon}z_h^{2\epsilon}\times\nn\\
    &&\sum_{I_1}\mathrm{Tr}\left[\overline{\mathcal{M}}_{q\to q,r}^{[\nu}\,(\bar{k},\bar{p},q)\,\left([\slash{\bar{p}},\gamma^\rho]-\tfrac{2\,w}{v\,Q^2}[\slash{\bar{p}},\slash{\bar{k}}]\,\bar{p}^\rho\right)\,\mathcal{M}^{\mu];\sigma}_{q\to qg,r}(\bar{k},\bar{p},\zeta\,\bar{p},q)\,[\slash{\bar{k}},\slash{S}]\gamma_5]\right]+\mathrm{c.c.}\,.\label{eq:Wantiq2qgr}
\end{eqnarray}
Those dynamical contributions need to be combined with the intrinsic and kinematical fragmentation contributions in \eqref{eq:Wq2qgvInt}, \eqref{eq:Wq2qgrInt}. Before we discuss this point, let us first calculate the LO contribution from \eqref{eq:Wsymq2qgv} and \eqref{eq:Wantiq2qgv} in a straight-forward manner. As in \eqref{eq:WLOkinsym}, \eqref{eq:WLOkinanti}, we simply have $\overline{\mathcal{M}}^\nu_{q\to q,\mathrm{LO}}=\gamma^\nu$. The leading order amplitude for the quark-gluon fragmentation reads:
\begin{equation}
    \mathcal{M}^{\mu;\sigma}_{q\to qg,\mathrm{LO}} = -\frac{i}{(1-\zeta)\,Q^2}\,[\gamma^\mu (\slash{\bar{k}}_{\mathrm{Br}}-(1-\zeta)\,\slash{\bar{p}}_{\mathrm{Br}})\gamma^\sigma]\,.\label{eq:Mq2qgLO}
\end{equation}
Inserting \eqref{eq:Mq2qgLO} into the hadronic tensors \eqref{eq:Wsymq2qgv}, \eqref{eq:Wantiq2qgv}, along with $\overline{\mathcal{M}}^\nu_{q\to q,\mathrm{LO}}=\gamma^\nu$, leads to the dynamical contributions to the symmetric and antisymmetric hadronic tensors at LO. We find,
\begin{eqnarray}
    \tilde{W}^{\mu\nu}_{\mathrm{LO,q\to qg,\,sym}}(S) & = & \dd z_h\,\left[\tfrac{8\,M_h\,x_B}{Q^4}\,(x_B\,P+q)^{\{\mu}\,\epsilon^{\nu\}PqS}\right]\sum_q e_q^2\,h_1^q(x_B)\,\left((1-\epsilon)\int_0^1 \dd \zeta\,\frac{\mathrm{Im}\hat{H}^{qg}_{FU}(z_h,\zeta)}{1-\zeta}\right) \,,\label{eq:WLOdynsym}\\
    \tilde{W}^{\mu\nu}_{\mathrm{LO,q\to qg,\,anti}}(S) & = & \dd z_h\,\left[\tfrac{8\,i\,M_h\,x_B}{Q^4}\,(x_B\,P+q)^{[\mu}\,\epsilon^{\nu ]PqS}\right]\sum_q e_q^2\,h_1^q(x_B)\,\left((1-\epsilon)\int_0^1 \dd \zeta\,\frac{\mathrm{Re}\hat{H}^{qg}_{FU}(z_h,\zeta)}{1-\zeta}\right) \,.\label{eq:WLOdynanti}
\end{eqnarray}
We observe that the dynamical contributions to the LO hadronic tensor, \eqref{eq:WLOdynsym} and \eqref{eq:WLOdynanti}, do not satisfy electromagnetic current conservation $q_\mu \tilde{W}^{\mu\nu}=q_\nu \tilde{W}^{\mu\nu}=0$, and hence, \eqref{eq:WLOdynsym} and \eqref{eq:WLOdynanti} alone are meaningless.

\subsubsection{Equation  of motion relations\label{sub:EoM}}

In order to re-establish the electromagnetic current conservation of the nucleon spin-dependent hadronic tensor, intrinsic, kinematical, and dynamical twist-3 contributions need to be combined. This can be achieved through the application of the QCD equation of motion (EoM). As shown in Ref.~\cite{Kanazawa:2015ajw}, one can use the QCD equation of motion to relate the various twist-3 fragmentation functions. The following EoM relations are relevant for this work,
\begin{eqnarray}
    (1-\epsilon)\int_0^1\dd\zeta\,\frac{\mathrm{Im}\hat{H}^{qg}_{FU}(z,\zeta)}{1-\zeta} & = & \frac{H^q(z)}{2\,z}+(1-\epsilon)\,H_1^{\perp (1),q}(z)\,,\label{eq:EoMIm}\\
    (1-\epsilon)\int_0^1\dd\zeta\,\frac{\mathrm{Re}\hat{H}^{qg}_{FU}(z,\zeta)}{1-\zeta} & = & -\frac{E^q(z)}{2\,z}\,.\label{eq:EoMRe}  
\end{eqnarray}
Note that we modified the relations of \cite{Kanazawa:2015ajw} in \eqref{eq:EoMIm}, \eqref{eq:EoMRe} to hold in arbitrary $d=4-2\epsilon$ dimensions for vanishing quark masses $m_q=0$. Eqs.~\eqref{eq:EoMIm}, \eqref{eq:EoMRe} are expected to hold for regularized yet unrenormalized fragmentation functions. Interestingly, it is explicitly stated in Ref.~\cite{Ma:2017upj} that the EoM relations \eqref{eq:EoMIm}, \eqref{eq:EoMRe} are also satisfied by the UV singularities of the involved fragmentation functions. Hence, we can expect that the EoM relations hold for the $\overline{\mathrm{MS}}$-renormalized fragmentation functions as well. We will come back to this feature later. \\

At this point, let us proceed with combining the LO symmetric tensors \eqref{eq:WLOkinsym}, \eqref{eq:WLOdynsym} and the antisymmetric LO tensors \eqref{eq:WLOkinanti}, \eqref{eq:WLOdynanti} using the EoM relations \eqref{eq:EoMIm}, \eqref{eq:EoMRe}. In this way, the nucleon spin-dependent symmetric hadronic tensor at LO accuracy reads,
\begin{eqnarray}
    \tilde{W}^{\mu\nu}_{\mathrm{LO,sym}}(S) & = &\tilde{W}^{\mu\nu}_{\mathrm{LO,q\to q,\,sym}}(S)+\tilde{W}^{\mu\nu}_{\mathrm{LO,q\to qg,\,sym}}(S)\nn\\
    & = &\dd z_h\, \left[\tfrac{2}{Q^3}\,\epsilon^{PqS\{\mu}(2\,x_B\,P+q)^{\nu\}}\right]\,\left[-\tfrac{4\,M_h}{Q}\sum_q e_q^2\,x_B\,h_1^q(x_B)\,\left(\tfrac{H^q(z_h)}{2\,z_h}+(1-\epsilon)\,H_1^{\perp (1),q}(z_h)\right)\right]\nn\\
    &\equiv &\dd z_h\,w^{\mu\nu}_\mathrm{s}\,F_{UT,\mathrm{LO}}^{\sin\phi_s}(x_B,z_h,Q)\,.\label{eq:WsymLO}
\end{eqnarray}
Note that the symmetric tensorial structure $w^{\mu\nu}_\mathrm{s}=\tfrac{2}{Q^3}\,\epsilon^{PqS\{\mu}(2\,x_B\,P+q)^{\nu\}}$ does satisfy electromagnetic current conservation $q_\mu w^{\mu\nu}_\mathrm{s}=q_\nu w^{\mu\nu}_\mathrm{s}=0$. It is the QCD equation of motion that restores this important feature. Similar observations have been reported in earlier works, e.g., Refs.~\cite{Gamberg:2018fwy,Schlegel:2012ve,Mulders:1995dh,Bacchetta:2006tn}. In the last line of \eqref{eq:WsymLO}, we introduced the notation $F_{UT,\mathrm{LO}}^{\sin\phi_s}(x_B,z_h,Q)$ to match our result to a model-independent structure function decomposition of the spin-dependent $\bm{P}_{h\perp}$-integrated SIDIS cross section reported in Ref.~\cite{Bacchetta:2006tn}. Interestingly, according to Ref.~\cite{Bacchetta:2006tn}, at LO, the structure function $F_{UT}^{\sin\phi_s}(x_B,z_h,Q)$ receives contributions only from twist-3 fragmentation functions, but not from twist-3 correlations in the nucleon. In this sense, the formula for the LO structure function,
\begin{eqnarray}
    F_{UT,\mathrm{LO}}^{\sin\phi_s}(x_B,z_h,Q) & = & -\tfrac{4\,M_h}{Q}\sum_q e_q^2\,x_B\,h_1^q(x_B)\,\left(\tfrac{H^q(z_h)}{2\,z_h}+(1-\epsilon)\,H_1^{\perp (1),q}(z_h)\right)\nn\\
    & \overset{\mathrm{EoM}}{=} & -\tfrac{4\,M_h}{Q}\sum_q e_q^2\,x_B\,h_1^q(x_B)\,\left((1-\epsilon)\int_0^1 \dd \zeta\,\tfrac{\mathrm{Im}\hat{H}^{qg}_{FU}(z_h,\zeta)}{1-\zeta}\right)\,,\label{eq:FUTLO}
\end{eqnarray}
can be considered complete.

The situation is somewhat different for the LO contribution to the antisymmetric part of the hadronic tensor,
\begin{eqnarray}
    \tilde{W}^{\mu\nu}_{\mathrm{LO,anti,frag}}(S) & = &\tilde{W}^{\mu\nu}_{\mathrm{LO,q\to q,\,anti}}(S)+\tilde{W}^{\mu\nu}_{\mathrm{LO,q\to qg,\,anti}}(S)\nn\\
    & = &\dd z_h\, \left[\tfrac{2\,i}{Q^3}\,\epsilon^{PqS[\mu}(2\,x_B\,P+q)^{\nu]}\right]\,\left[-\tfrac{4\,M_h}{Q}\sum_q e_q^2\,x_B\,h_1^q(x_B)\,\left(\tfrac{E^q(z_h)}{2\,z_h}\right)\right]\nn\\
    &\equiv &\dd z_h\,w^{\mu\nu}_\mathrm{a}\,F_{LT,\mathrm{LO,frag}}^{\cos(\phi_s)}(x_B,z_h,Q)\,.\label{eq:WantiLO}
\end{eqnarray}
As for the symmetric tensor, the antisymmetric structure $w^{\mu\nu}_\mathrm{a}=\tfrac{2\,i}{Q^3}\,\epsilon^{PqS[\mu}(2\,x_B\,P+q)^{\nu]}$ also satisfies electromagnetic current conservation. The corresponding LO structure function in \eqref{eq:WantiLO}, again matched to the structure function decomposition of Ref.~\cite{Bacchetta:2006tn}, reads,
\begin{eqnarray}
    F_{LT,\mathrm{LO,frag}}^{\cos\phi_s}(x_B,z_h,Q) & = & -\tfrac{4\,M_h}{Q}\sum_q e_q^2\,x_B\,h_1^q(x_B)\,\tfrac{E^q(z_h)}{2\,z_h}\nn\\
    & \overset{\mathrm{EoM}}{=} & +\tfrac{4\,M_h}{Q}\sum_q e_q^2\,x_B\,h_1^q(x_B)\,\left((1-\epsilon)\int_0^1 \dd \zeta\,\frac{\mathrm{Re}\hat{H}^{qg}_{FU}(z_h,\zeta)}{1-\zeta}\right)\,.\label{eq:FLTLO}
\end{eqnarray}
However, \eqref{eq:FLTLO} does not constitute the complete LO formula. According to \cite{Bacchetta:2006tn}, twist-3 correlations in the nucleon may also contribute to the structure function $F_{LT}^{\cos\phi_s}$. As mentioned  before, we will disregard these contributions and focus on twist-3 fragmentation effects only.

\subsubsection{Quark-Antiquark-Gluon correlations\label{sub:qqg}}
\begin{figure}
    \centering
     \includegraphics[width=0.6\linewidth]{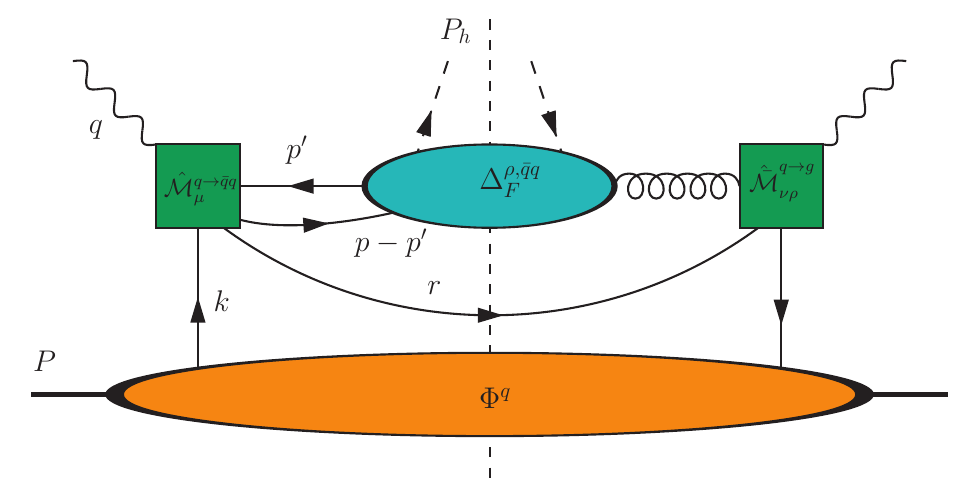}
    \caption{Collinear twist-3 factorization setup: dynamical quark-quark-gluon fragmentation.}
    \label{fig:3Partonqqg}
\end{figure}
There is yet another dynamical twist-3 configuration of partons fragmenting into a hadron that we haven't considered so far. These contributions are depicted in Fig.~\ref{fig:3Partonqqg}. They consist of an amplitude with a \textit{quark-antiquark} pair fragmenting together into a hadron, interfering with an amplitude that has a fragmenting gluon. Such a configuration may be called \textit{antiquark-quark-gluon} contributions. It does not appear at LO but at NLO. The factorization ansatz sketched in Fig.~\ref{fig:3Partonqqg} can be directly translated into a formula for the hadronic tensor, as discussed in the previous subsection. Analogous to \eqref{eq:Wq2qgrInt}, we obtain
\begin{eqnarray}
    \tilde{W}^{\mu\nu}_{q\to \bar{q}q,r}
    &=&\dd z_h\,\int_{x_B}^1 \tfrac{\dd w}{w}\int_{z_h}^1\tfrac{\dd v}{v}\,\int_0^1 \dd \zeta\,(-ig_{\perp,\rho\sigma})\,\tfrac{x_B}{2\,w}\,\int \tfrac{\dd^{d-2}\bm{P}_{h\perp}}{(2\pi)^{d-1}}\,\delta\left(\tfrac{1-v}{v}\tfrac{1-w}{w}\,z_h^2Q^2-\bm{P}_{h\perp}^2\right)\times\nn\\
    &&\sum_qe_q^2\,\sum_{I_1}\mathrm{Tr}\left[\overline{\mathcal{M}}_{q\to g,r}^{\nu;\sigma}\,(\bar{k},\bar{p},q)\,\left(\Delta_F^{\bar{q}q;\rho}(\tfrac{z_h}{v},\zeta)\,\mathcal{M}^{\mu}_{q\to \bar{q}q,r}(\bar{k},\bar{p},\zeta\,\bar{p},q)\right)\,\Phi^q(\tfrac{x_B}{w})\right]+\mathrm{c.c.}\,.\label{eq:Wq2qqrInt}
\end{eqnarray}
The non-perturbative matrix element $\Delta^{\bar{q}q}$ that appears in \eqref{eq:Wq2qqrInt} has the form of $\Delta^{qg}$ in \eqref{eq:Deltaqgq}, but with a reversed role of the quark and the gluon field (cf.~\cite{Gamberg:2018fwy}),
\begin{eqnarray}
\Delta^{\bar{q}q;\rho}_{F;ij}(z,\zeta) & = &   \frac{1}{N_c}\sumint\int_{-\infty}^\infty\tfrac{\dd \lambda}{2\pi}\int_{-\infty}^\infty\tfrac{\dd \mu}{2\pi}\,\e^{-i\tfrac{\lambda}{z}\zeta}\e^{-i\tfrac{\mu}{z}(1-\zeta)}\langle0|\,[\infty\, n; 0]\,F^{n\rho}(0)\,[0;\infty\,n]\,|P_h X\rangle\times\nn\\
&&\hspace{4cm}\langle P_hX|\,[\infty\,n;\lambda\,n]\,q_j(\lambda\,n)\,ig\mu_R^\epsilon\,\bar{q}_j(\mu\,n)\,[\mu\,n;\infty\,n]\,|0\rangle   \nn\\
&= & M_h\,z^{-1+2\epsilon}\left[\tfrac{i}{2}\left([\slash{P}_h,\gamma^\rho]-\tfrac{2x_B}{z_h\,Q^2}[\slash{P}_h,\slash{P}]\,P_h^\rho \right)\,\left(\mathrm{Im}\hat{H}^{\bar{q}q}_{FU}(z,\zeta)+i\,\mathrm{Re}\hat{H}^{\bar{q}q}_{FU}(z,\zeta)\right)+...\right]\,.\label{eq:Deltaqqg}
\end{eqnarray}
Note that the dynamical fragmentation functions $\hat{H}^{qg}_{FU}$ and $\hat{H}^{\bar{q}q}_{FU}$, despite their similar definitions \eqref{eq:Deltaqgq} and \eqref{eq:Deltaqqg}, are not related to each other and must be considered independent functions. Inserting the parameterizations \eqref{eq:Deltaqqg} and \eqref{eq:Phi} into \eqref{eq:Wq2qqrInt} leads to a similar separation into a symmetric and antisymmetric part, as shown in \eqref{eq:Wsymq2qgr} and \eqref{eq:Wantiq2qgr}. We refrain from writing this separation explicitly here. 

We also note that the role of the fragmenting quark-antiquark pair is symmetric under the exchange of the quark and the antiquark. Instead of a matrix element $\Delta^{\bar{q}q}$, we could have worked with a reversed matrix element $\Delta^{q\bar{q}}$ in \eqref{eq:Wq2qqrInt} (with an interchange of the quark fields $q\leftrightarrow \bar{q}$; see also \cite{Gamberg:2018fwy}) and parameterized it as in \eqref{eq:Deltaqqg} in terms of fragmentation functions $\hat{H}^{q\bar{q}}_{FU}$. As argued in \cite{Gamberg:2018fwy}, these two fragmentation functions are related,
\begin{equation}
    \hat{H}_{FU}^{q\bar{q}}(z,\zeta)= \hat{H}_{FU}^{\bar{q}q}(z,1-\zeta)\,.\label{eq:CparHqq}
\end{equation}
Since we integrate over the full support $0\le \zeta\le 1$ in \eqref{eq:Wq2qqrInt}, we conclude that both types of fragmentation functions of \eqref{eq:CparHqq} are implicitly included in this formula.

\subsubsection{Renormalization of twist-3 fragmentation functions\label{sub:Renorm}}
The last key ingredient vital for the completion of the NLO calculation is the renormalization of the hadronic matrix elements in the LO formulae \eqref{eq:WsymLO} and \eqref{eq:WantiLO}. Up to this point, we have considered \textit{bare} parton distributions and twist-3 fragmentation functions that contain explicit UV-divergences, which can be regularized in $d=4-2\epsilon$ dimensions. On the other hand, the hard partonic scattering parts contain collinear divergences, as discussed in \eqref{eq:epsExp}. If a QCD factorization theorem has been established and proven for a given observable, then one can expect that the UV-divergences of the hadronic matrix elements cancel the collinear divergences of the hard part in perturbation theory, order by order. If, however, a full, all-order QCD factorization theorem is not available for a given observable (such as the twist-3 observables studied in this paper), one can at least test the QCD factorization to order $\mathcal{O}(\alpha_s)$ if the UV-divergences of the hadronic matrix elements are known. Typically, the UV-divergences of an ordinary twist-2 parton distribution/ fragmentation functions are determined by the DGLAP evolution kernels, sometimes also called the splitting functions.

For example, our LO expressions \eqref{eq:WsymLO} and \eqref{eq:WantiLO} both contain the quark \textit{transversity} distribution $h_1^q(x)$ \cite{Ralston:1979ys,Artru:1989zv,Jaffe:1991kp,Jaffe:1991ra,Cortes:1991ja}. The LO splitting function of the transversity distribution was known early on (see, e.g., \cite{Vogelsang:1997ak}), as were the evolution kernels and UV-poles. In the $\overline{\mathrm{MS}}$-scheme, we can therefore provide the relation between the bare transversity distribution and the $\overline{\mathrm{MS}}$-renormalized (and finite) transversity distribution as follows,
\begin{eqnarray}
    h_{1,\mathrm{bare}}^q(x) & = & h_{1}^{q,\overline{\mathrm{MS}}}(x,\mu)+\frac{\alpha_s}{2\pi}\frac{S_\epsilon}{\epsilon}\int_{x}^1\tfrac{\dd w}{w}\,(\Delta_TP_{qq})(w)\,h_{1}^{q,\overline{\mathrm{MS}}}(\tfrac{x}{w},\mu)+\mathcal{O}(\alpha_s^2)\,,\label{eq:TransversityMSbar}\\
    (\Delta_TP_{qq})(w) & = & C_F\left[\frac{2\,w}{(1-w)_+}+\frac{3}{2}\delta(1-w)\right]\,.\label{eq:Transvsplittingfunction}
\end{eqnarray} 
Note that the $\overline{\mathrm{MS}}$-renormalized transversity distribution $h_{1}^{q,\overline{\mathrm{MS}}}$ depends on the factorization scale $\mu$. The quantity $\Delta_TP_{qq}$ in \eqref{eq:Transvsplittingfunction} is the leading order splitting function for the quark transversity distribution. We again encounter the quantity $S_\epsilon\equiv (4\pi)^\epsilon/\Gamma(1-\epsilon)$, which is in line with the $\overline{\mathrm{MS}}$-scheme at one loop order. Also, note that no mixing to a corresponding transversity distribution for a gluon appears in \eqref{eq:TransversityMSbar}. In fact, there exists no collinear gluon transversity distribution.

In the case of the collinear twist-3 fragmentation functions included in Eqs.~\eqref{eq:WsymLO}, \eqref{eq:WantiLO}, \eqref{eq:FUTLO}, and \eqref{eq:FLTLO}, we can again use results obtained in the literature. In Ref.~\cite{Ma:2017upj}, the LO evolution kernels (or splitting functions) were calculated for the chiral-odd intrinsic twist-3 fragmentation functions $H^q(z)$ and $E^q(z)$, for the chiral-odd kinematical twist-3 fragmentation function $H_1^{\perp(1),q}(z)$, and for the dynamical twist-3 fragmentation functions $\hat{H}_{FU}^{qg}(z,\zeta)$ and $\hat{H}_{FU}^{\bar{q}q}(z,\zeta)$. However, for this paper, only the splitting functions for those fragmentation functions that appear in the LO formulae \eqref{eq:WsymLO} and \eqref{eq:WantiLO} are relevant. To this end, we simply read off the results of Ref.~\cite{Ma:2017upj} and rewrite them using our notation. Eventually, this gives us a prescription for the $\overline{\mathrm{MS}}$-renormalization. We obtain,
\begin{eqnarray}
    \left(\frac{H^q(z)}{2\,z}+(1-\epsilon)\,H_1^{\perp (1),q}(z)\right)\Big|_{\mathrm{bare}} & = & \frac{H^{q,\overline{\mathrm{MS}}}(z,\mu)}{2\,z}+(1-\epsilon)\,H_1^{\perp (1),q,\overline{\mathrm{MS}}}(z,\mu) \label{eq:SFsym}\\
    &&\hspace{-4cm}+ \frac{\alpha_s}{2\pi}\frac{S_\epsilon}{\epsilon} \int_z^1 \tfrac{\dd v}{v}\int_0^1\dd \zeta\,\left[\Delta\hat{P}_{qg,\mathrm{Im}}(v,\zeta)\,\mathrm{Im}[\hat{H}_{FU}^{qg}(\tfrac{z}{v},\zeta)]+\Delta\hat{P}_{\bar{q}q,\mathrm{Im}}(v,\zeta)\,\mathrm{Im}[\hat{H}_{FU}^{\bar{q}q}(\tfrac{z}{v},\zeta)]\right]+\mathcal{O}(\alpha_s^2)\,,\nn\\
     \left(-\frac{E^q(z)}{2\,z}\right)\Big|_{\mathrm{bare}} & = & -\frac{E^{q,\overline{\mathrm{MS}}}(z,\mu)}{2\,z} \label{eq:SFanti}\\
    &&\hspace{-4cm}+ \frac{\alpha_s}{2\pi}\frac{S_\epsilon}{\epsilon} \int_z^1 \tfrac{\dd v}{v}\int_0^1\dd \zeta\,\left[\Delta\hat{P}_{qg,\mathrm{Re}}(v,\zeta)\,\mathrm{Re}[\hat{H}_{FU}^{qg}(\tfrac{z}{v},\zeta)]+\Delta\hat{P}_{\bar{q}q,,\mathrm{Re}}(v,\zeta)\,\mathrm{Re}[\hat{H}_{FU}^{\bar{q}q}(\tfrac{z}{v},\zeta)]\right]+\mathcal{O}(\alpha_s^2)\,.\nn
\end{eqnarray}
The explicit form of the four evolution kernels reads \cite{Ma:2017upj},
\begin{eqnarray}
    \Delta\hat{P}_{qg,\mathrm{Im}}(v,\zeta) & = & C_F\,\left[\frac{-2}{(1-v)_+}\frac{v-\zeta}{(1-\zeta)^2}-\frac{1+\zeta}{\zeta(1-\zeta)^2}-\frac{2v}{\zeta}+\frac{2+\zeta}{\zeta(1-\zeta)}-\frac{\delta(1-v)}{2\,(1-\zeta)}\left(1-\frac{2\,\ln\zeta}{1-\zeta}\right)\right]\label{eq:SFqgIm}\\
    &&+N_c\,\left[\frac{1}{2\,\zeta\,(1-\zeta)}-\frac{1-v}{\zeta(1-\zeta)^2\,(1-v\,\zeta)}-\delta(1-v)\,\frac{\ln\zeta}{2\,(1-\zeta)^2}\right]\,,\nn
\end{eqnarray}
\begin{eqnarray}
    \Delta\hat{P}_{\bar{q}q,\mathrm{Im}}(v,\zeta) & = & \frac{1}{2\,N_c}\left[\frac{(1-v)(1-2v)}{\zeta\,(1-v+v\,\zeta)}\right]\,,\label{eq:SFqbqIm}
\end{eqnarray}
\begin{eqnarray}
    \Delta\hat{P}_{qg,\mathrm{Re}}(v,\zeta) & = & C_F\,\left[\frac{-2}{(1-v)_+}\frac{v}{1-\zeta}-\frac{1}{\zeta}-\frac{\delta(1-v)}{2\,(1-\zeta)}\left(1-\frac{2\,\ln\zeta}{1-\zeta}\right)\right]\label{eq:SFqgRe}\\
    &&+N_c\,\left[\frac{1}{2\,(1-\zeta)}+\frac{1}{2\,\zeta}-\delta(1-v)\,\frac{\ln\zeta}{2\,(1-\zeta)^2}\right]\,,\nn
\end{eqnarray}
\begin{eqnarray}
    \Delta\hat{P}_{\bar{q}q,\mathrm{Re}}(v,\zeta) & = & \frac{1}{2\,N_c}\left[\frac{1-v}{\zeta\,(1-v+v\,\zeta)}\right]\,.\label{eq:SFqbqRe}
\end{eqnarray}
We find that the insertion of the explicit expressions \eqref{eq:SFsym}, \eqref{eq:SFanti} for the bare fragmentation functions into the LO formulae \eqref{eq:FUTLO} and \eqref{eq:FLTLO} eventually removes all collinear poles in the complete NLO calculation and provides a well-defined NLO formula in terms of $\overline{\mathrm{MS}}$-renormalized transversity distributions and twist-3 fragmentation functions. This feature, together with the fact that the splitting functions \eqref{eq:SFqgIm} - \eqref{eq:SFqbqRe} originate from a completely independent calculation \cite{Ma:2017upj}, provides strong support for the validity of the collinear twist-3 factorization. We note that Ref.~\cite{Ma:2017upj} indicates that all of the LO evolution kernels satisfy the equation-of-motion relations \eqref{eq:EoMIm} and \eqref{eq:EoMRe}. This immediately implies that the $\overline{\mathrm{MS}}$-renormalized fragmentation functions also satisfy these relations, and we may replace all renormalized intrinsic and kinematical twist-3 fragmentation functions at LO in favor of renormalized dynamical ones, as indicated in the last lines of \eqref{eq:FUTLO} and \eqref{eq:FLTLO}.

\subsection{NLO calculation\label{sub:NLO}}
\begin{figure}
    \centering
    \includegraphics[width=0.5\linewidth]{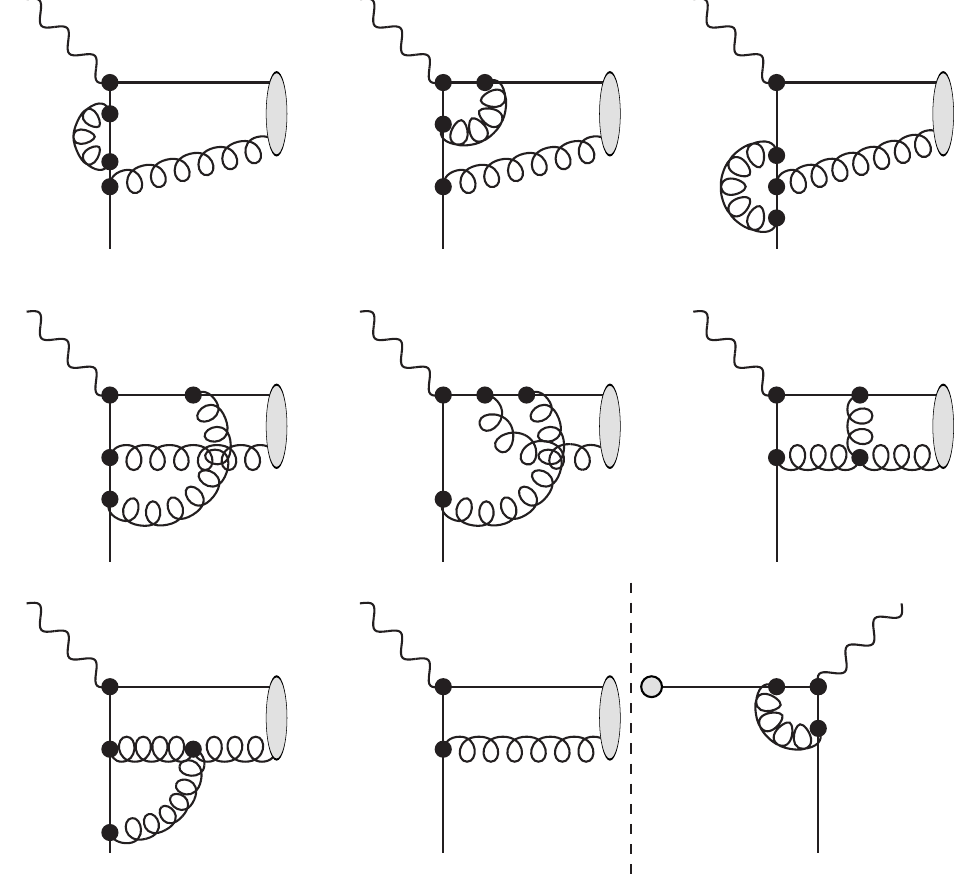}
    \caption{Virtual graphs contributing to dynamical twist-3 $qg$ fragmentation.}
    \label{fig:virtNLOdyn}
\end{figure}
We are now in a position to discuss the NLO corrections to the LO results of the structure functions \eqref{eq:FUTLO} and \eqref{eq:FLTLO}. We start with the virtual corrections. They contribute to the intrinsic/kinematical twist-3 $q\to q$ fragmentation channel, as well as to the dynamical twist-3 $q\to qg$ fragmentation channel. Since the general factorization formulae were already established in \eqref{eq:Wq2qgvInt} (intrinsic/kinematical) and in \eqref{eq:Wsymq2qgv}, \eqref{eq:Wantiq2qgv} (dynamical), we need to calculate the one-loop correction to the partonic amplitudes $\mathcal{M}^\mu_{q\to q,v}$ and $\mathcal{M}^{\mu;\sigma}_{q\to qg,v}$. The one-loop correction to $\mathcal{M}^\mu_{q\to q,v}$ is simply the usual QCD vertex correction of the quark-photon vertex. However, the one-loop QCD correction to $\mathcal{M}^{\mu;\sigma}_{q\to qg,v}$ is somewhat more complicated and is provided by several loop diagrams shown in Fig.~\ref{fig:virtNLOdyn}. We calculate these loop diagrams in the light-cone gauge \eqref{eq:LCgauge}. The procedure for how we perform the loop integrals has been discussed at length in  Refs.~\cite{Gamberg:2018fwy,Rein:2025qhe}. We do not wish to repeat these explanations here and refer the reader to these references. We do note, however, that we observe one important feature that also shows up in Refs.~\cite{Gamberg:2018fwy,Rein:2025qhe}: The sum of the contributions depicted in Fig.~\ref{fig:virtNLOdyn} is independent of the gauge parameter $\kappa$ in \eqref{eq:LCprop}. Hence, we would obtain the same result for a calculation performed directly in Feynman gauge. This important feature supports our result.
\\

\begin{figure}
    \centering
    \includegraphics[width=0.5\linewidth]{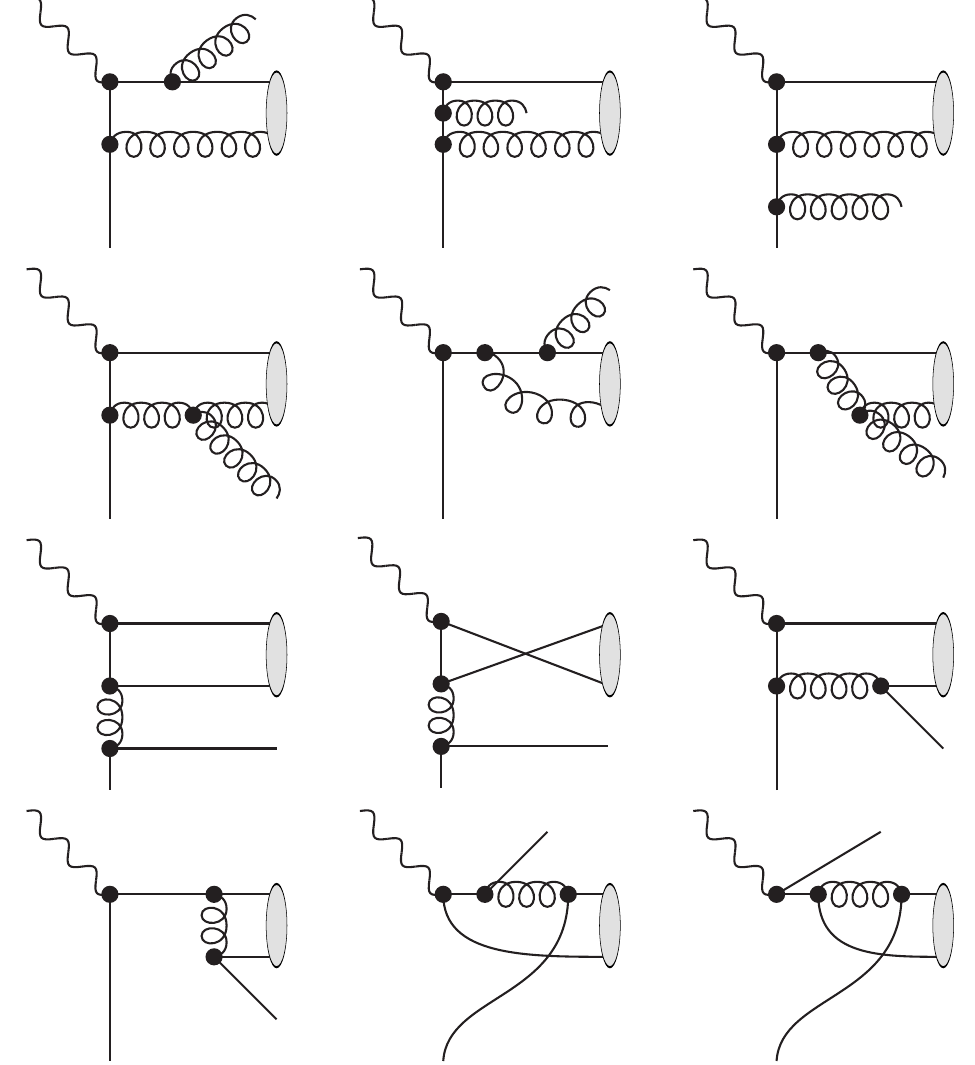}
    \caption{Real graphs contributing to dynamical twist-3 $qg$ and $\bar{q}q$ fragmentation.}
    \label{fig:realNLOdyn}
\end{figure}
In the next step, we discuss the real corrections with one unobserved parton in the final state. The factorization formulae for this situation have already been set up in \eqref{eq:Wrealunpol}, \eqref{eq:Wrealsym}, \eqref{eq:Wrealanti} for the kinematical twist-3 fragmentation, and in \eqref{eq:Wsymq2qgr}, \eqref{eq:Wantiq2qgr}, \eqref{eq:Wq2qqrInt} for the dynamical contributions. With those formulae at hand, we only need to calculate the real QCD emission corrections at order  $\mathcal{O}(\alpha_s)$. As an example, we show the diagrams with real gluon/quark emissions in the final state for the dynamical contributions in Fig.~\ref{fig:realNLOdyn}. In this figure, the first six diagrams contribute to the dynamical $q\to gq$-channel, while the last six contribute to the $q\to \bar{q}q$-channel discussed above. Of course, these contributions need to be interfered with the corresponding real emission diagrams of the $q\to q$-channel and $q\to g$-channel (not shown here). It is straight-forward to calculate these contributions using the formulae \eqref{eq:Wrealunpol}, \eqref{eq:Wrealsym}, \eqref{eq:Wrealanti}, \eqref{eq:Wsymq2qgr}, \eqref{eq:Wantiq2qgr}, \eqref{eq:Wq2qqrInt}. The only phase space integrals that need to be performed are those in \eqref{eq:PhpInts}. We note that for the real corrections, the dynamical contributions and the intrinsic/kinematical contributions are neither color nor e.m. gauge invariant by themselves. Both types of contributions need to be combined using the equation-of-motion relations \eqref{eq:EoMIm}, \eqref{eq:EoMRe}. After this procedure, we observe that the gauge parameter $\kappa$ drops out, and the final result will be proportional to the symmetric and antisymmetric tensorial structures $w_s^{\mu\nu}$, $w_a^{\mu\nu}$ in \eqref{eq:WsymLO}, \eqref{eq:WantiLO} that respect e.m. current conservation. As before, we consider this behavior a strong check for our calculation.
\\

The final test of the validity of the collinear twist-3 factorization approach for fragmentation in $\bm{P}_{h\perp}$-integrated SIDIS at NLO is the cancellation of the $1/\epsilon^2$ and $1/\epsilon$-poles that appear in both the virtual and real corrections after expansion in $\epsilon$. As mentioned before, these poles appear in dimensional regularization and can be interpreted as soft and collinear divergences. We first note that the $1/\epsilon^2$ pole of the virtual loop diagrams and the one generated by the real diagrams (cf. \eqref{eq:epsExp}) cancel each other. This cancellation is expected, as stated by the Kinoshita-Lee-Nauenberg theorem \cite{Kinoshita:1962ur,Lee:1964is}. The remaining collinear $1/\epsilon$-pole is more complicated. In order to explicitly observe the cancellation of this type of divergence, we need to include the $\overline{\mathrm{MS}}$-renormalization of the hadronic matrix elements \eqref{eq:TransversityMSbar}, \eqref{eq:SFsym} and \eqref{eq:SFanti}, as discussed in the previous section. In this way, we eventually obtain finite hard scattering coefficients at the level of the hadronic tensor for both symmetric and antisymmetric parts.

\section{Analytical results\label{sec:Analytics}}

In this section, we present our analytical results for the spin independent $\bm{P}_{h\perp}$-integrated SIDIS cross section at NLO in pQCD, as well as the twist-3 fragmentation contributions to the transverse nucleon spin-dependent cross section. The latter constitutes the main result of this paper and is $\--$ to the best of our knowledge $\--$ new. The former, i.e., the unpolarized cross section, was computed to NLO accuracy about 40 years ago (see, e.g., \cite{Furmanski:1981cw,deFlorian:1997zj}). We re-evaluated the unpolarized cross section as a test of our factorization approach and found exact agreement with Ref.~\cite{deFlorian:1997zj}.

\subsection{Unpolarized cross section}
We present the NLO result for the unpolarized cross section in the following way: 

First, we complete the calculation of the hadronic tensors \eqref{eq:q2qvInt} and \eqref{eq:q2qrInt} for unpolarized nucleons and unpolarized detected hadrons, evaluating them to NLO accuracy. After $\overline{\mathrm{MS}}$-renormalization of the unpolarized PDF $f_1^q$ and FF $D_1^q$, we obtain a finite and well-defined result for the partonic cross section of the $q\to q$-channel. There are two more partonic channels at NLO that also need to be calculated, with a gluon acting as a parton in the nucleon ($g\to q$-channel) or acting as a fragmenting parton ($q\to g$-channel). Since this is a standard feature, we do not further discuss gluon PDFs/FFs in this paper and refer the reader to Ref.~\cite{deFlorian:1997zj} instead.

Then we contract the unpolarized hadronic tensor, evaluated at NLO accuracy, with the leptonic tensor \eqref{eq:leptTens} and insert the result into the cross section formula \eqref{eq:SIDIScs}. Finally, the cross section is expressed as $\dd \sigma/\dd x_B \dd y \dd z_h$. A comparison with a structure function decomposition of this cross section (see Ref.~\cite{Bacchetta:2006tn}) allows for the identification of the unpolarized structure functions $F_{UU,T/L}$ by means of the following formula\footnote{Target mass corrections are not taken into account in this paper.}, 
\begin{eqnarray}
    \frac{\dd \sigma_{UU}}{\dd x_B \,\dd y \,\dd z_h}&=&\frac{4\pi \alpha_{\rm em}^2}{ x_B\,y\, Q^2}\left[ \left(1-y+\frac{y^2}{2}\right)F_{UU,T} +(1-y)F_{UU,L}\right].\label{eq:CSunpSF}
\end{eqnarray}
While the structure function $F_{UU,T}$ receives contributions within the pQCD approach already at leading-order (LO) accuracy, i.e., the parton model, via \eqref{eq:WLOunpol}, the structure function $F_{UU,L}$ is populated by pQCD contributions starting at the next-to-leading-order (NLO) level.

Our results for the two unpolarized structure functions take the following form:
\begin{eqnarray}
    F_{UU,T}(x_B,z_h,Q)&=& x_B\sum_{q,\bar{q}} e_q^2 \Big\{f_1^q(x_B,\mu^2)D_1^q(z_h,\mu^2)\label{eq:FUUTNLO}\\
        &&+ \frac{\alpha_s}{2\pi}\left[f_1^q \otimes \mathcal{C}_1^{q \to q} \otimes D_1^q +f_1^g \otimes \mathcal{C}_{1}^{g \to q} \otimes D_1^q+f_1^q \otimes \mathcal{C}_1^{q \to g} \otimes D_1^g\right](x_B,z_h,\mu^2) \Big\}+\mathcal{O}(\alpha_s^2)\,,\nn
    \end{eqnarray}    
\begin{eqnarray}        
    F_{UU,L}(x_B,z_h,Q)&=&x_B \frac{\alpha_s}{2\pi}\sum_{q,\bar{q}} e_q^2 \Big[f_1^q \otimes \mathcal{C}_L^{q \to q} \otimes D_1^q  +f_1^g \otimes \mathcal{C}_L^{g \to q} \otimes D_1^q+f_1^q \otimes \mathcal{C}_L^{q \to g} \otimes D_1^g \Big](x_B,z_h,\mu^2)\label{eq:FUULNLO}   \\
    &&+\mathcal{O}(\alpha_s^2)\,,\nn
\end{eqnarray}
where we introduced the notation for the double convolution 
\begin{equation}
    f \otimes \mathcal{C} \otimes D \equiv \int_{x_B}^1 \tfrac{\dd w}{w}\int_{z_h}^1 \tfrac{\dd v}{v} \,f\left(\tfrac{x_B}{w},\mu^2\right) \mathcal{C}\left(w,v,\tfrac{Q^2}{\mu^2}\right) D\left(\tfrac{z_h}{v},\mu^2\right)\,.\label{eq:DoubleConv}
\end{equation}
The explicit form of the partonic factors $\mathcal{C}_1$ and $\mathcal{C}_L$ for the three partonic subprocesses $q\to q$, $g\to q$, and $q\to g$ can be found in Appendix \ref{appendix:unpolarized}. As mentioned before, they agree with the results of Ref.~\cite{deFlorian:1997zj}.

\subsection{Unpolarized lepton, transversely polarized nucleon\label{sub:UT}}
We proceed with the presentation of the structure function $F_{UT}^{\sin\phi_s}$ of the $\bm{P}_{h\perp}$-integrated SIDIS cross section for an unpolarized lepton and a transversely polarized nucleon, as introduced in \eqref{eq:FUTLO} and calculated at LO accuracy. In the following, we present the same structure function calculated to NLO accuracy within the collinear twist-3 formalism. The details were discussed in the previous section. Our explicit analytic result $\--$ valid in $d=4$ dimensions $\--$ reads:
\begin{eqnarray}
        F_{UT,\mathrm{NLO}}^{\sin\phi_s}(x_B,z_h,Q)&=& -\frac{4 M_h}{Q}\sum_{q,\bar{q}} e_q^2 \Bigg\{x_B\,h_1^q(x_B,\mu^2)\,\int_0^1 \dd \zeta\,\frac{\mathrm{Im}\hat{H}_{FU}^{qg}(z_h,\zeta,\mu^2)}{1-\zeta}\nn\\
      &  &+ \frac{\alpha_s}{2\pi}\int_{x_B}^1\tfrac{\dd w}{w}\int_{z_h}^1\tfrac{\dd v}{v}\int_0^1\dd \zeta\,\,\tfrac{x_B}{w}\,h_1^q \left(\tfrac{x_B}{w},\mu^2\right)\nn\\
    &    & \times \Bigg[\mathcal{C}_{UT}^{q \to qg}(w,v,\zeta,\tfrac{Q^2}{\mu^2}) \frac{\mathrm{Im} \hat{H}_{FU}^{qg}\left(\frac{z_h}{v},\zeta,\mu^2\right)}{1-\zeta}+\mathcal{C}_{UT}^{q \to \bar{q} q}(w,v,\zeta,\tfrac{Q^2}{\mu^2})\, \mathrm{Im} \hat{H}_{FU}^{\bar{q}q}\left(\tfrac{z_h}{v},\zeta,\mu^2\right)\Bigg]\Bigg\}\nn\\
    && + \mathcal{O}(\alpha_s^2)\,,\label{eq:F_UT@NLO}
\end{eqnarray}
where the explicit forms of the coefficient functions $\mathcal{C}_{UT}$ are shown in Eqs.~\eqref{eq:CUTcolorsplit} and \eqref{eq:CUTq2qq}. As mentioned before, we only consider NLO corrections to the chiral-odd twist-3 fragmentation effects in this paper. The results of Ref.~\cite{Bacchetta:2006tn} indicate that at LO, there are no contributions to the structure function $F_{UT}^{\sin\phi_s}$ from twist-3 effects that originate from the transversely polarized nucleon. However, it may happen that contributions from twist-3 effects in the nucleon appear at the NLO level. Due to the absence at LO, such potential contributions would necessarily need to be finite and well-behaved on their own. In other words, any occurrence of collinear singularities in the hard partonic cross sections in twist-3 effects within the polarized nucleon would immediately violate the collinear twist-3 factorization. 

Interestingly, Ref.~\cite{Zhang:2025fvt} reports an NLO calculation of twist-3 effects in the transversely polarized nucleon, but for the single-spin asymmetry in the transverse momentum-integrated Drell-Yan process. It was found in Ref.~\cite{Zhang:2025fvt} that collinear twist-3 factorization holds at the NLO level for this particular observable. However, this DY observable, in contrast to $F_{UT}^{\sin\phi_s}$ in SIDIS, does receive a leading order contribution from the twist-3 effect in the polarized nucleon. 

Whether or not twist-3 effects in the nucleon emerge at the NLO level for the $\bm{P}_{h\perp}$-integrated SIDIS SSA is not clear at the moment, and we plan to investigate this question in future work. At this point, we assume that Eq.~\eqref{eq:F_UT@NLO} constitutes the complete NLO result.

\subsection{Longitudinally polarized lepton, transversely polarized nucleon}
In addition, we present our results for the NLO corrections to the twist-3 fragmentation contributions to the doubly polarized structure function $F_{LT}^{\cos\phi_s}$ in Eq.~\eqref{eq:FLTLO}. Our explicit analytical result in $d=4$ dimensions reads:
\begin{eqnarray}
    F_{LT,\mathrm{NLO},\mathrm{frag}}^{\cos\phi_s}(x_B,z_h,Q) & = & \frac{4 M_h}{Q}\sum_{q,\bar{q}} e_q^2 \Bigg\{
        x_B\,h_1^q(x_B,\mu^2)\,\int_0^1\dd\zeta\,\frac{\mathrm{Re}\hat{H}_{FU}^{qg}(z_h,\zeta,\mu^2)}{1-\zeta}\nn\\
        &  &+ \frac{\alpha_s}{2\pi}\int_{x_B}^1\tfrac{\dd w}{w}\int_{z_h}^1\tfrac{\dd v}{v}\int_0^1\dd \zeta\,\,\tfrac{x_B}{w}\,h_1^q \left(\tfrac{x_B}{w},\mu^2\right)\nn\\
    &    & \times \Bigg[\mathcal{C}_{LT}^{q \to qg}(w,v,\zeta,\tfrac{Q^2}{\mu^2}) \,\frac{\mathrm{Re} \hat{H}_{FU}^{qg}\left(\frac{z_h}{v},\zeta,\mu^2\right)}{1-\zeta}+\mathcal{C}_{LT}^{q \to \bar{q} q}(w,v,\zeta,\tfrac{Q^2}{\mu^2})\, \mathrm{Re} \hat{H}_{FU}^{\bar{q}q}\left(\tfrac{z_h}{v},\zeta,\mu^2\right)\Bigg]\Bigg\}\nn\\
    && + \mathcal{O}(\alpha_s^2)\,,\label{eq:F_LT@NLO}
\end{eqnarray}
where the explicit forms of the coefficient functions $\mathcal{C}_{LT}$ can be found in Eqs.~\eqref{eq:CLTcolorsplit} and \eqref{eq:CLTq2qq}. Unlike our result for the structure function $F_{UT}^{\sin\phi_s}$ \eqref{eq:F_UT@NLO}, we do not consider \eqref{eq:F_LT@NLO} a complete NLO result, since twist-3 effects within the transversely polarized nucleon (already present at LO) are not taken into account in this paper. We plan to investigate these effects in the future.

\section{Numerical predictions\label{sec:Numerics}}
This section is devoted to testing the numerical implications of our NLO formula \eqref{eq:F_UT@NLO} for the structure function $F_{UT}^{\sin\phi_s}(x_B,z_h,Q)$ of an unpolarized lepton colliding with a transversely polarized nucleon. To this end, we focus on semi-inclusive charged pion production in polarized electron-proton collisions $ep^\uparrow\to e\pi^{\pm} X$. This particular process has been measured by the HERMES collaboration, and experimental data have been released for the transverse nucleon single-spin asymmetry (SSA) $A_{UT}^{\sin\phi_s}$ \cite{HERMES:2020ifk}. We perform an exploratory numerical study of the SSA $A_{UT}^{\sin\phi_S}$ by introducing certain model scenarios for the otherwise unknown dynamical twist-3 fragmentation functions $\mathrm{Im}\hat{H}_{FU}^{\pi/qg}(z,\zeta)$ and $\mathrm{Im}\hat{H}_{FU}^{\pi/\bar{q}q}(z,\zeta)$. Eventually, we find that a comparison to the HERMES data allows us to exclude some of the model scenarios for the fragmentation functions $\mathrm{Im}\hat{H}_{FU}^{\pi/qg}(z,\zeta)$ and $\mathrm{Im}\hat{H}_{FU}^{\pi/\bar{q}q}(z,\zeta)$ $\--$ a feature that is absent in a LO pQCD analysis but emerges at NLO.\\

We adjust our numerical approach to the kinematics of the HERMES experiment \cite{HERMES:2020ifk}. In particular, we assume a center-of-mass (c.m.) energy $\sqrt{s}\approx7\,\rm{GeV}$. However, due to the limited acceptance of the HERMES experiment, a direct one-to-one comparison of the theoretical $\bm{P}_{\pi\perp}$-integrated cross section/ structure function \eqref{eq:F_UT@NLO} to the HERMES data is difficult. This feature has been discussed in detail in the JAM-3D LO analysis of this observable \cite{Gamberg:2022kdb}. The complication of the comparison \textit{theory vs.~data} arises because the HERMES experiment detected pions (and kaons) in a limited range of transverse momentum $\bm{P}_{\pi\perp}$. Strictly speaking, the effects measured at HERMES were not \textit{fully} $\bm{P}_{\pi\perp}$-integrated. On the other hand, HERMES released so-called \textit{$x_B$- and $z_h$-projections} of their data. For this reason, it was suggested in Ref.~\cite{Gamberg:2022kdb} to use these projections as a proxy for the fully $\bm{P}_{\pi \perp}$-integrated observable. Indeed, Ref.~\cite{Gamberg:2022kdb} directly compared the HERMES $x_B$ and $z_h$-projections to the following observable, 
\begin{equation}
    \tilde{A}_{UT}^{\sin\phi_s}\equiv\frac{F_{UT}^{\sin\phi_s}(x_B,z_h,Q)}{F_{UU,T}(x_B,z_h,Q)+f(y)\,F_{UU,L}(x_B,z_h,Q)}\,,\quad\mathrm{with}\,\,f(y)=\frac{1-y}{1-y+\frac{y^2}{2}}\,\,\mathrm{and}\,\,y=\frac{Q^2}{x_B\,s}\,,\label{eq:AUT}
\end{equation}
where the LO formulae \eqref{eq:FUTLO}, as well as Eqs.~\eqref{eq:FUUTNLO}, \eqref{eq:FUULNLO} (with $\alpha_s\to 0$), were used as input in \eqref{eq:AUT} for the polarized and unpolarized structure functions $F_{UT}^{\sin\phi_s}$ and $F_{UU,T}$, $F_{UU,L}$, respectively. Note that the structure function $F_{UU,L}$ vanishes at LO.  In addition, the averaged $x_B$- and $z_h$-values of each experimental bin were inserted in the observable $A_{UT}^{\sin\phi_s}$ in \eqref{eq:AUT}. 

\subsection{Scenarios for the dynamical twist-3 fragmentation functions $\mathrm{Im}\hat{H}_{FU}$}

In our exploratory study, we follow the procedure of the JAM-3D LO analysis but insert the NLO results \eqref{eq:F_UT@NLO}, along with \eqref{eq:FUUTNLO}, \eqref{eq:FUULNLO}, instead of the LO results. This approach requires input for the chiral-odd dynamical twist-3 fragmentation functions $\mathrm{Im}\hat{H}_{FU}^{\pi/qg}(z,\zeta)$ and $\mathrm{Im}\hat{H}_{FU}^{\pi/\bar{q}q}(z,\zeta)$ \textit{on their full support} $0<z<1$ and $0<\zeta<1$. Currently, only very limited information on these functions is available in the literature.

However, the LO JAM-3D analysis \cite{Gamberg:2022kdb} extracted the  twist-3 fragmentation functions (favored and disfavored),
\begin{equation}
    \frac{\tilde{H}^q(z,\mu)}{2z}\equiv \frac{H^q(z,\mu)}{2z}+H_1^{\perp (1),q}(z,\mu) \overset{\mathrm{EoM}}{=} \int_0^1\dd\zeta\,\frac{\mathrm{Im}\hat{H}_{FU}^{qg}(z,\zeta,\mu)}{1-\zeta}\,,\label{eq:Htilde}
\end{equation}
together with the transversity PDF $h_1^q$ from the HERMES data \cite{HERMES:2020ifk}. We can utilize this result to construct three scenarios for the full dynamical twist-3 fragmentation functions in the following way: We assume the following model ansatz for the functions of interest while simultaneously respecting all of the few known constraints,
\begin{eqnarray}   
        \mathrm{Im} \hat{H}^{qg}_{FU}(z,\zeta,\mu) &= &\frac{\tilde{H}^q(z,\mu)}{2 z}\,\mathcal{F}^q\left(\zeta;\,N_q,a_q,b_q,\gamma_q,\tilde{a}_q,\tilde{b}_q\right)\label{eq:ImHFUqgAnsatz}\,,\\
        \mathrm{Im} \hat{H}^{\bar{q}q}_{FU}(z,\zeta,\mu) &=& \frac{\tilde{H}^q(z,\mu)}{2 z}\,\mathcal{F}^q\left(\zeta;\,\bar{N}_q,c_q,c_q,\bar{\gamma}_q,\tilde{c}_q,\tilde{c}_q\right).\label{eq:ImHFUqbqAnsatz}
\end{eqnarray}
We use the LO JAM-3D extraction of $\tilde{H}$ as input in the ansatz \eqref{eq:ImHFUqgAnsatz}. Consequently, our functions $\mathrm{Im}\hat{H}_{FU}$ in \eqref{eq:ImHFUqgAnsatz} and \eqref{eq:ImHFUqbqAnsatz} inherit the dependence on the factorization scale $\mu$ from the LO JAM-3D extraction of $\tilde{H}$. In Ref.~\cite{Gamberg:2022kdb}, a modified DGLAP evolution was assumed for the fragmentation function $\tilde{H}$, which may, at best, be considered an approximation of the true twist-3 evolution governed by the evolution kernels in Eqs.~\eqref{eq:SFsym}. Only recently has a numerical code, the so-called Honeycomb/Snowflake code \cite{Rodini:2024usc}, been released that numerically evolves quark-gluon-quark correlation functions in the nucleon based on the true twist-3 LO evolution. However, a corresponding code for twist-3 fragmentation functions is not available at the moment, so we consider applying the modified DGLAP evolution of the JAM-3D analysis to be the best option we currently have. While we plan to investigate the twist-3 evolution of fragmentation matrix elements in the future, this requires a dedicated study and is beyond the scope of this paper.

Furthermore, we introduced a phenomenological function $\mathcal{F}$ in the model ansatz \eqref{eq:ImHFUqgAnsatz}. Note that the ansatz \eqref{eq:ImHFUqgAnsatz} is chosen such that the $z$- and $\zeta$-dependence of the twist-3 fragmentation functions $\mathrm{Im}\hat{H}_{FU}$ factorizes. The phenomenological function $\mathcal{F}$ then models the dependence on the variable $\zeta$. We choose a functional form that is often used in the literature, in particular in the JAM-3D analysis \cite{Gamberg:2022kdb},
    
\begin{table}[]
    \centering
    \begin{tabular}{c| c c c c c c c c c c | c c c c c c c c c c c}
           $ $ & \multicolumn{10}{c}{$qg$}&\multicolumn{8}{c}{$\bar{q}q$}\\
          &$a_u$&$a_d$&$b_u$&$b_d$&$\tilde a_u$&$\tilde a_d$&$\tilde b_u$&$\tilde b_d$&$\gamma_u$&$\gamma_d$&$\bar N_u$&$\bar N_d$&$c_u$&$c_d$& $\tilde c_u$&$\tilde c_d$& $\bar \gamma_u$ &$\bar \gamma_d$ \\
          \hline
         $S_1$&$1.51$&$1.25$&$1.86$&$1.89$&$0.36$&$0.95$& $0.17$& $0.33$&   $-1.54$&$-2.49$&$1.14$&$1.06$&$2.30$&$1.48$& $0.31$&$0.76$& $-1.57$ &$-1.88$ \\
         $S_2$&$1.55$&$1.85$&$1.36$&$1.65$&$0.51$&$0.18$&$0.01$&$0.16$&$-1.47$&$-3.32$&$1.58$&$0.61$&$1.38$&$1.26$& $0.15$&$0.69$& $-2.24$ &$-3.44$ \\
         $S_3$&$2.46$&$2.49$&$1.32$&$1.91$&$0.29$&$ 0.39$&$0.55$&$0.72$&$-3.64$&$-1.97$&$1.49$&$ 2.81$&$2.05$&$1.76$& $0.93$&$0.13$& $-2.26$ &$ -2.59$ \\
         \hline\hline
    \end{tabular}
    \caption{Model parameters for the scenarios $S_1$, $S_2$ and $S_3$.}
    \label{tab:scenarios}
\end{table}

\begin{equation}
    \mathcal{F}^q\left(\xi;\,N_q,a_q,b_q,\gamma_q,\tilde{a}_q,\tilde{b}_q\right) \equiv \frac{N_q\, \xi^{a_q} (1-\xi)^{b_q}\left( 1+\gamma_q\, \xi^{\tilde{a}_q}(1-\xi)^{\tilde{b}_q}\right)}{\mathbb{B}(a_q+2,b_q+1) + \gamma_q\, \mathbb{B}(a_q+\tilde{a}_q+2 , b_q + \tilde{b}_q +1)},\label{eq:FAnsatz}
\end{equation} 
where $\mathbb{B}(a,b)=\Gamma(a)\Gamma(b)/\Gamma(a+b)$ is the Euler beta function, and $\xi$ is a generic fractional momentum variable. 
It is important to note that, for the factorized ansatz of the fragmentation function $\mathrm{Im}\hat{H}_{FU}^{qg}$ in \eqref{eq:ImHFUqgAnsatz}, our model is guaranteed to satisfy the EoM relation \eqref{eq:EoMIm} by construction. One can readily verify this statement from the functional form \eqref{eq:FAnsatz} by noting that the following relation,
\begin{equation}
    \int_0^1 \dd \zeta\, \frac{\mathcal{F}(\zeta)}{1-\zeta}=1\,,\label{eq:IntFEoM}
\end{equation}
holds, provided we set $N_q=1$. 
In addition, there are also model-independent boundary conditions for both the dynamical twist-3 functions $\hat{H}_{FU}^{qg}$ and $\hat{H}_{FU}^{\bar{q}q}$, which have been derived in Refs.~\cite{Meissner:2008yf,Kanazawa:2015ajw,Gamberg:2018fwy}. According to these papers, not only do the fragmentation functions $\hat{H}_{FU}^{qg}$ and $\hat{H}_{FU}^{\bar{q}q}$ vanish at the boundaries $\zeta=0$ and $\zeta=1$, but also their derivatives. Hence, we expect $\Im \hat{H}^{qg}_{FU}(z,0)=\Im \hat{H}^{qg}_{FU}(z,1)= \partial_\zeta \Im \hat{H}^{qg}_{FU}(z,\zeta)|_{\zeta\to 0}=\partial_\zeta \Im \hat{H}^{qg}_{FU}(z,\zeta)|_{\zeta\to 1}=0$. These boundary constraints are easily translated into constraints for the phenomenological function $\mathcal{F}$, 
\begin{eqnarray}
    & \mathcal{F}(0)=\mathcal{F}(1)=0\,, & \nn\\
    & \left.\pdv{\mathcal{F}(\zeta)}{\zeta}\right\rvert_{\zeta\to0}=\left.\pdv{\mathcal{F}(\zeta)}{\zeta}\right\rvert_{\zeta\to1}=0\,. &\label{eq:BoundaryFAnsatz}
\end{eqnarray}
We adjust the model parameters in \eqref{eq:FAnsatz} so that the conditions in \eqref{eq:BoundaryFAnsatz} are met.

We note that the dynamical fragmentation function $\hat{H}_{FU}^{\bar{q}q}$ does not appear in constraint equations like the EoM-relation \eqref{eq:EoMIm}. For this reason, there is a priori no need for the phenomenological function $\mathcal{F}$ in \eqref{eq:ImHFUqbqAnsatz} to satisfy the constraint \eqref{eq:IntFEoM}. We chose to model the fragmentation function $\mathrm{Im}\hat{H}_{FU}^{\bar{q}q}$ in \eqref{eq:ImHFUqbqAnsatz} to be as similar as possible to the quark-gluon counterpart $\mathrm{Im}\hat{H}_{FU}^{qg}$ in \eqref{eq:ImHFUqgAnsatz}.
As discussed below Eq.~\eqref{eq:CparHqq}, we expect the $\bar{q}q$ correlator to be symmetric under the exchange of quarks and antiquarks, and consequently under the exchange of $\zeta\leftrightarrow1-\zeta$. Hence, we set the exponents of $\zeta$ and $(1-\zeta)$ to be equal for a given flavor $q$ for this reason.

We proceed by specifying scenarios for the fragmentation functions provided by the ansätze \eqref{eq:ImHFUqgAnsatz} and \eqref{eq:ImHFUqbqAnsatz}. 
We define a scenario as a specific set of model parameters, including both fragmentation functions. This means that the $k$-th scenario is described by a specific set $S_k=\{a_u,a_{\bar{u}},a_d,a_{\bar{d}},b_u,b_{\bar{u}},\dots \}$. Among these models, we select three scenarios that are particularly useful for our purpose, i.e., to demonstrate that our NLO analysis, in principle, allows us to significantly constrain the dynamical functions $\mathrm{Im}\hat{H}_{FU}$ over the full support, which is otherwise impossible with an LO analysis. To be specific, for scenarios 1, 2, and 3, we chose the sets of model parameters specified in Tab.~\ref{tab:scenarios}.

\subsection{Numerical predictions for HERMES and a future EIC\label{sub:HERMESEIC}}
\begin{figure}[]
    \centering
    \includegraphics[width=0.9\linewidth]{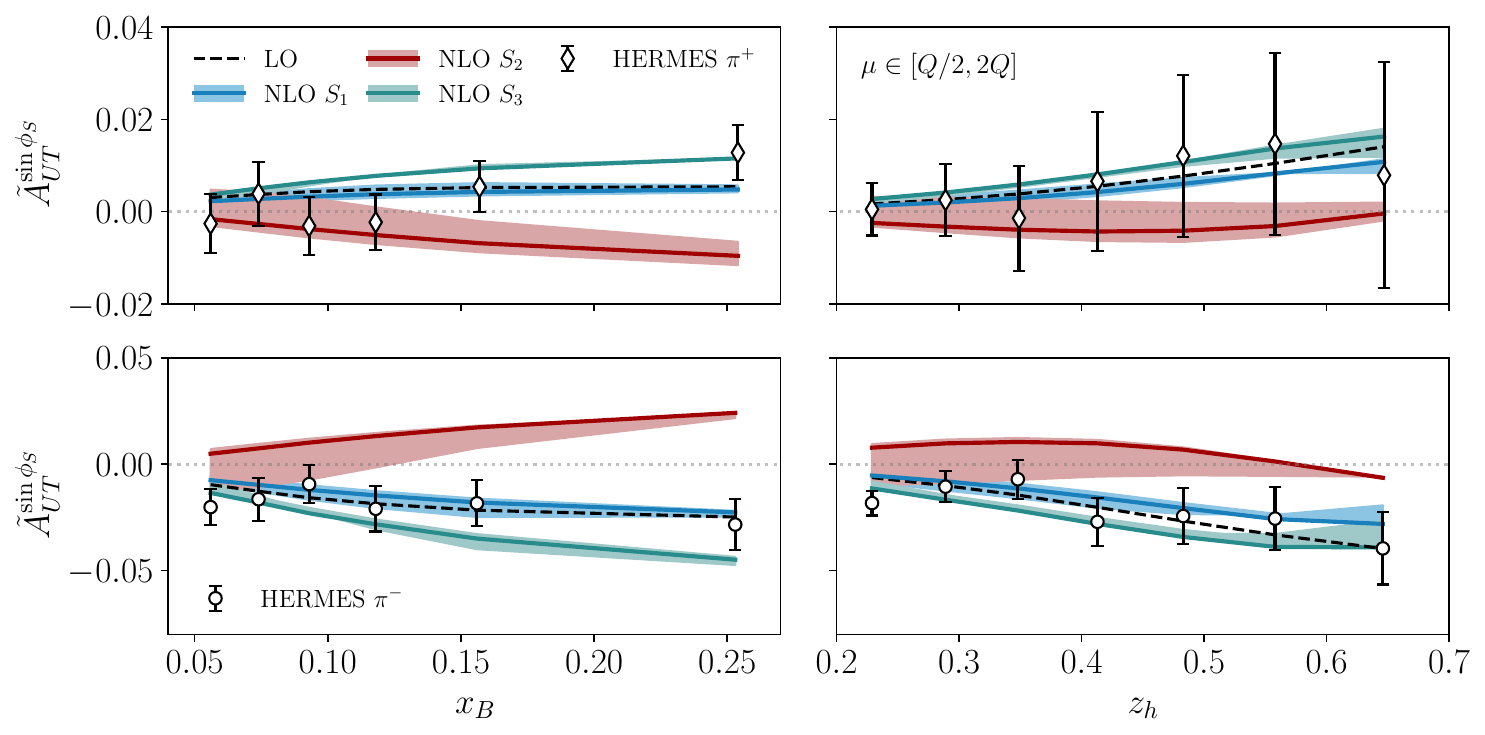}
    \caption{The $x_B$ and $z_h$ projections of the $A_{UT}^{\sin\phi_s}$ asymmetry, computed at NLO accuracy, against HERMES data. Three different model scenarios $S_k$ for the $qg$ and $\bar{q}q$ fragmentation functions are shown. Shaded regions represent the scale variation $\mu\in[Q/2,2Q]$.}
    \label{fig:HERMES_A_UT_NLO}
\end{figure}

The results of our exploratory NLO numerical analysis are presented in Fig.~\ref{fig:HERMES_A_UT_NLO}. In these plots, our theoretical NLO predictions are shown based on the asymmetry \eqref{eq:AUT} with the NLO structure functions $F_{UT,\mathrm{NLO}}^{\sin\phi_s}$ from Eq.~\eqref{eq:F_UT@NLO}, and $F_{UU,T} $, $F_{UU,L}$ from Eqs.~\eqref{eq:FUUTNLO}, \eqref{eq:FUULNLO}, along with the scenarios in table \ref{tab:scenarios} for the dynamical fragmentation functions. 
For the remaining twist-2 PDFs and FFs that enter the factorization formulae \eqref{eq:F_UT@NLO} and \eqref{eq:FUUTNLO}, \eqref{eq:FUULNLO}, we used existing extractions from data fits available in the literature. To be specific, for the transversity distribution, we used the LO JAM-3D extraction of Ref.~\cite{Gamberg:2022kdb}. For the unpolarized PDF $f_1^{q/g}$ of the proton, we used the CJ15 sets \cite{Accardi:2016qay}, and for the pion FF $D_1^{q/g}$, we used the parameterization \cite{deFlorian:2014xna}.

Our NLO curves are compared with the HERMES \cite{HERMES:2020ifk} projections along $x_B$ and $z_h$ of the single-spin asymmetry in Fig.~\ref{fig:HERMES_A_UT_NLO} for both $\pi^+$ and $\pi^-$ production. 

The NLO curves corresponding to the three different scenarios in table \ref{tab:scenarios} exhibit quite different behaviors. The scenario $S_1$ yields very similar curves compared to the LO case and describes the data qualitatively well. This is not an indication that the NLO correction is small in general, but rather that the $\mathcal{O}(\alpha_s)$ correction of the numerator of the SSA is roughly equal in size and opposite in sign compared to the $\mathcal{O}(\alpha_s)$ correction originating from the denominator. Scenario $S_2$ instead showcases significant corrections all the way up to $\sim 50-100\%$. In fact, NLO corrections are so large that the asymmetry changes sign compared to the LO case. Lastly, scenario $S_3$ shows a similar trend to the LO case while somewhat describing an enhanced asymmetry effect. The analysis of these three scenarios strongly indicates that the NLO corrections to the asymmetry are quite sensitive to the choice of parameters used to describe the dynamical fragmentation functions, as well as their overall model parametrization. In principle, one should perform a fit to the experimental data to constrain the dynamical fragmentation functions. This is, however, beyond the scope of this work and may be the subject of future investigations.
We would like to point out that the LO curves in Fig.~\ref{fig:HERMES_A_UT_NLO} best agree with the HERMES data. This is not at all surprising, since the LO input for the transversity function $h_1^q$ and the twist-3 fragmentation function $\tilde{H}^q$, taken from the JAM-3D fit \cite{Gamberg:2022kdb}, was designed to agree with the data. 

At last, we note that the theoretical error bands of the curves in Fig.~\ref{fig:HERMES_A_UT_NLO} originate from a variation of the factorization scale between $\mu=Q/2$ and $\mu=2\,Q$. As mentioned before, the observed scale variations do not originate from the full twist-3 evolution of the fragmentation function of $\mathrm{Im}\hat{H}_{FU}$, but from a modified DGLAP evolution of the function $\tilde{H}$ inherited through the ansatz \eqref{eq:ImHFUqgAnsatz} from Ref.~\cite{Gamberg:2022kdb}.

\begin{figure}[]
    \centering
    \includegraphics[width=0.9\linewidth]{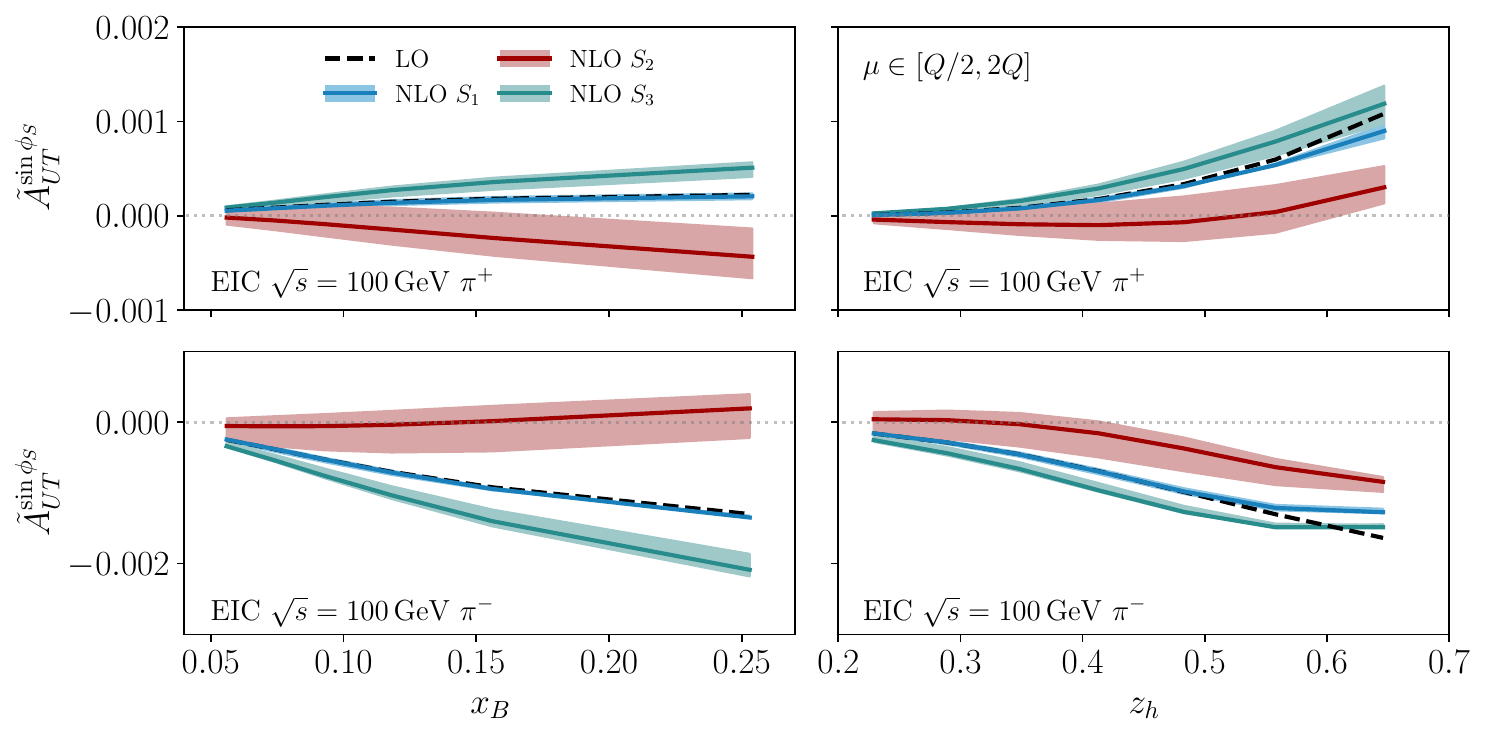}
    \caption{The $x_B$ and $z_h$ projections of the $A_{UT}^{\sin\phi_s}$ asymmetry at NLO for EIC kinematics. Three different model scenarios $S_k$ for the $qg$ and $\bar{q}q$ fragmentation functions are shown. Shaded regions represent the scale variation $\mu\in[Q/2,2Q]$.}
    \label{fig:EIC_A_UT_NLO}
\end{figure}

Eventually, we can prepare an NLO prediction for the transverse single-nucleon spin asymmetry in $\bm{P}_{h\perp}$-integrated SIDIS at an EIC. In other words, we study the NLO curves for kinematics relevant to the future EIC. In order to do so, we take the kinematical points ${x_B,y,z_h}$ similar to the HERMES data but assumed a higher c.m. energy of $\sqrt{s}\approx 100\,\rm{GeV}$. This immediately implies that the virtuality of the exchanged photon is larger since it is fixed by the SIDIS relation $Q^2=s \,x_B \,y$. The resulting curves are presented in  Fig.~\ref{fig:EIC_A_UT_NLO}. Concerning the different scenarios in table \ref{tab:scenarios}, we observe analogous behavior compared to the HERMES case: $S_1$ is similar to LO, $S_2$ leads to a sign change of the asymmetry compared to LO, and $S_3$ enhances the LO effect. The overall asymmetry at an EIC is roughly a factor of 10 smaller compared to the HERMES experiment, consistent with the fact that we are studying a sub-leading twist, $1/Q$-suppressed observable.  

There are essentially two important outcomes of this exploratory NLO study. First, we see that the NLO corrections to the numerator of the SSA can be large and are not at all negligible. In scenario $S_2$, we even observe NLO corrections so large that the asymmetry changes sign. This fact alone should be enough to emphasize the importance of NLO corrections to spin-dependent observables in QCD. The fact that NLO corrections to the numerator of SSAs can be large has been numerically tested in a very similar manner for single inclusive hadron production $ep\to hX$ \cite{Rein:2025pwu}. Secondly, the fact that the NLO single-spin asymmetry shows great sensitivity to the set of employed twist-3 matrix elements further motivates the importance of future precision colliders, such as the EIC. In fact, future accurate data collected at larger center-of-mass energies will deepen our understanding of sub-leading twist distribution and fragmentation effects, shedding light on detailed hadron structure and the hadronization mechanism.

\section{Conclusions \& Outlook\label{sec:Conclusions}}

In this paper, we performed a pQCD analysis of the $\bm{P}_{h\perp}$-integrated SIDIS cross section for a transversely polarized nucleon at NLO accuracy. We particularly focused on contributions from chiral-odd multi-parton contributions in the fragmentation process. We discussed in detail the setup of the collinear twist-3 factorization for the symmetric and antisymmetric hadronic tensor. Most importantly, we explicitly observe that all collinear singularities cancel at the one-loop level after $\overline{\mathrm{MS}}$-renormalization of the transversity distribution and twist-3 fragmentation functions. 
We find that collinear factorization holds for both the symmetric and anti-symmetric hadronic tensor. They are related to different structure functions/ observables in $\bm{P}_{h\perp}$-integrated SIDIS, the SSA for a transversely polarized incident nucleon and an unpolarized lepton, and a double-spin asymmetry for a transversely polarized nucleon and a longitudinally polarized lepton. Both observables may receive contributions at NLO from multi-parton correlations in the polarized nucleon as well. Although we did not investigate such contributions in this paper, we plan to do so in future work.

By assuming somewhat realistic scenarios/ models for the chiral-odd twist-3 fragmentation functions, we numerically computed our NLO results for the transverse SSA and confronted them with actual data taken from the HERMES experiment. While some of the scenarios can very well describe the data, other scenarios are ruled out by the data. We presented NLO predictions for the SSA  at an EIC, with the hope of stimulating future experimental measurements.
In the future, we plan to work towards an actual data fit for the chiral-odd multi-parton fragmentation functions based on the NLO results gathered in this paper. This would include the implementation of the proper twist-3 evolution for the fragmentation functions. While the LO evolution kernels are known \cite{Ma:2017upj}, a numerical code that performs the proper evolution is not available. It would be important for future endeavors to have such a code and we plan to work towards this goal.

\acknowledgments
We thank Werner Vogelsang and Daniel Pitonyak for valuable and helpful discussions.
This work has been supported by Deutsche Forschungsgemeinschaft (DFG) through the Research Unit FOR 2926 (Project No. 409651613).

\appendix
\section{Lorentz transformation and frames}\label{appendix:LT}
In this appendix, we provide some more details on the treatment of reference frames in our approach. As mentioned in Section \ref{sec:frames}, factorization is conveniently set up in the collinear hadron frame (CHF). As described in Eq.~\eqref{eq:collframe}, we have the following parameterization of the four-momenta $P^\mu$, $P_h^\mu$, $q^\mu$ in this frame,
\begin{equation}\label{eq:4vecCHF}
    \begin{aligned}
        P^\mu&\overset{\mathrm{CHF}}{=}\frac{Q}{2x_B}\Big(1,0,0,1\Big),\\
        P_h^\mu&\overset{\mathrm{CHF}}{=}\frac{z_hQ}{2}\Big(1,0,0,-1\Big),\\
        q^\mu&\overset{\mathrm{CHF}}{=}Q\Big(\frac{1}{2}\bm{\chi}_T^2,|\bm{\chi}_T|\cos\phi_\chi,|\bm{\chi}_T|\sin\phi_\chi,\frac{1}{2}\bm{\chi}_T^2-1\Big),\\
        q^\mu_T&\overset{\mathrm{CHF}}{=}Q\Big(0,|\bm{\chi}_T|\cos\phi_\chi,|\bm{\chi}_T|\sin\phi_\chi,0\Big),
    \end{aligned}
\end{equation}
where $\bm{\chi}_T\equiv \bm{q}_T/Q$, with $\bm{q_T}^2=-q_T^2$. We explicitly introduced the angle $\phi_\chi$ between the $\bm{\chi}_T$ vector and the $x$-axis. 
\\

The Lorentz transformation that converts the momenta \eqref{eq:4vecCHF}, expressed in the collinear hadron frame, to the Breit frame, can be readily constructed by various subsequent rotations and boosts. 
To be specific, we perform
\begin{itemize}
    \item a rotation $R_z(\phi_\chi)$ around the $z$-axis with angle $\phi_\chi$,
    \item a rotation $R_y(\psi)$ around the $y$-axis with angle $\psi$, such that $\cos\psi=\frac{1-\frac{1}{2}\bm{\chi}_T^2}{\sqrt{1+\frac{1}{4}\bm{\chi}_T^4}}$,
    \item a boost $B_z(\beta_z)$ in the $z$-direction with velocity $\beta_z=-\frac{\frac{1}{2}\bm{\chi}_T^2}{\sqrt{1+\frac{1}{4}\bm{\chi}_T^4}}$,
    \item a boost $B_x(\beta_x)$in the $x$-direction with velocity $\beta_x=\frac{|\bm{\chi}_T|}{1+\frac{1}{2}\bm{\chi}_T^2}$,
    \item an inverse rotation $R_z^{-1}(\phi_\chi)$ around the $z$-axis with angle $\phi_\chi$.
\end{itemize}
We emphasize that the angle $\psi$ satisfies the usual trigonometric property $-1\le\cos\psi\le1$ for any value of $\bm{\chi}_T^2$, and the same is true for $\sin\psi$. Similarly, both boost factors $\beta_z$ and $\beta_x$ are well-defined and satisfy $|\beta_{x/z}|<1$.
In particular, no restriction on the value of the transverse momentum $|\bm{q}_T|$ is required. The overall Lorentz transformation from the collinear hadron frame to the Breit frame is therefore given by:
\begin{equation}
    \Lambda^\mu_{\,\,\,\nu}=(R^{-1}_z(\phi_\chi))^\mu_{\,\,\,\rho} \,(B_x(\beta_x))^\rho_{\,\,\,\sigma}\, (B_z(\beta_z))^\sigma_{\,\,\,\eta} \, (R_y(\psi))^\eta_{\,\,\,\kappa} \, (R_z(\phi_\chi))^\kappa_{\,\,\,\nu}.\label{eq:LT}
\end{equation}
This Lorentz transformation can be expressed explicitly in matrix form,
\begin{equation}
   \Lambda^\mu_{\,\,\,\nu}=\left(\begin{array}{c c c c}
    1+\frac{1}{2}\bm{\chi}_T^2 & -|\bm{\chi}_T|\cos\phi_\chi & -|\bm{\chi}_T|\sin\phi_\chi & -\frac{1}{2}\bm{\chi}_T^2\\
    -|\bm{\chi}_T|\cos\phi_\chi  & 1 & 0 & |\bm{\chi}_T|\cos\phi_\chi \\
    -|\bm{\chi}_T|\sin\phi_\chi  & 0 & 1 & |\bm{\chi}_T|\sin\phi_\chi \\
    \frac{1}{2}\bm{\chi}_T^2 & -|\bm{\chi}_T|\cos\phi_\chi & -|\bm{\chi}_T|\sin\phi_\chi & 1-\frac{1}{2}\bm{\chi}_T^2
    \end{array}\right).\label{eq:LTmatrix}
\end{equation}
We note that this transformation satisfies all textbook requirements of a proper Lorentz transform, that is, $\Lambda^\mu_{\,\,\,\lambda}g_{\mu\nu}\Lambda^\nu_{\,\,\,\eta}=g_{\lambda \eta}$ or, in matrix notation, $\Lambda^T g \Lambda=g$, and $\det\Lambda=1$, for \textit{all} values of $\bm{q}_T$. Applying the transformation \eqref{eq:LTmatrix} to the vectors in \eqref{eq:4vecCHF} yields,
\begin{equation}\label{eq:4vecBF}
    \begin{aligned}
        P^\mu&\overset{\mathrm{BF}}{=}\frac{Q}{2x_B}\Big(1,0,0,1\Big),\\
        P_h^\mu&\overset{\mathrm{BF}}{=}\frac{z_hQ}{2}\Big(1+\bm{\chi}_T^2,-2\bm{\chi}_T,\bm{\chi}_T^2-1\Big),\\
        q^\mu&\overset{\mathrm{BF}}{=}\Big(0,0,0,-Q\Big),\\
        q^\mu_T&\overset{\mathrm{BF}}{=}Q\Big(-\bm{\chi}_T^2,\bm{\chi}_T,-\bm{\chi}_T^2\Big).
    \end{aligned}
\end{equation}
This corresponds to the form of the momenta \eqref{eq:Brframe}, \eqref{eq:PhBr} in the Breit frame after the identification $(\bm{q}_T)_{\rm CHF}=-\frac{1}{z_h}(\bm{P}_{h\perp})_{\rm BF}$, as discussed in Eq.~\eqref{eq:relPhpqT}. Again, we emphasize that the Lorentz transform \eqref{eq:LTmatrix} can always be performed and is not restricted by the specific kinematics of the SIDIS process. Therefore, the integration over the transverse hadron momentum $\bm{P}_{h\perp}$ in the Breit frame is not restricted either.

\section{Hard scattering coefficients}
\subsection{Unpolarized nucleon}
\label{appendix:unpolarized}
The hard scattering coefficients appearing in the unpolarized structure functions $F_{UU,T}$ and $F_{UU,L}$ in Eqs.~\eqref{eq:FUUTNLO}, \eqref{eq:FUULNLO} are given by
\begin{eqnarray}
          \mathcal{C}^{q\to q}_1 (v,w,\chi_Q)&=& C_F\,\Big(\left(3\ln\chi_Q-8\right)\,\delta(1-w)\delta(1-v)\nn\\
      &  &+\left[(1+v^2)\left(\frac{\ln(1-v)}{1-v}\right)_+ +1-v+(1+v^2)\frac{ \ln\chi_Q+\ln v}{(1-v)_+}\right]\delta(1-w)\nn\\
     &   &+\left[(1+w^2)\left(\frac{\ln(1-w)}{1-w}\right)_+ +1-w+(1+w^2)\frac{\ln\chi_Q-\ln w}{(1-w)_+}\right]\delta(1-v)\nn\\
    &&+\frac{2v^2w^2-2v^2w-2vw^2+4vw+v^2+w^2-2v-2w+2}{(1-w)_+\,(1-v)_+}\Big),\label{eq:C1q2q}
        \end{eqnarray}
\begin{eqnarray}
        \mathcal{C}^{g\to q}_1 (w,v,\chi_Q)&=&T_F\,\Big(\left[(w^2+(1-w)^2)\left(\ln\chi_Q+\ln \tfrac{1-w}{w}\right)+2w(1-w)\right]\delta(1-v)\nn\\
        && \hspace{5cm}\left.+\frac{(w^2+(1-w)^2)(v^2+(1-v)^2)}{v(1-v)_+}\right),\label{eq:C1g2q}\\
       \mathcal{C}^{q\to g}_1 (w,v,\chi_Q)&=& C_F\,\Big(\left[\tfrac{1+(1-v)^2}{v}\left(\ln\chi_Q+\ln \left(v(1-v)\right)\right)+v\right]\delta(1-w)\nn\\
       &&\hspace{5cm}+ \frac{1+v^2+w^2-2vw^2-2v^2w+2v^2w^2}{v(1-w)_+}\Big),\label{eq:C1q2g}
\end{eqnarray}
and
\begin{eqnarray}
          \mathcal{C}^{q\to q}_L (w,v)&=& C_F\,\left[4\, v\,w\right],\nn\\
        \mathcal{C}^{g\to q}_L (w,v)&=&T_F\,\left[8\,w\,(1-w)\right],\nn\\
         \mathcal{C}^{q\to g}_L (w,v)&=&C_F\left[4\,w\,(1-v)\right],\label{eq:CL}
\end{eqnarray}
where we used $\chi_Q\equiv \tfrac{Q^2}{\mu^2}$, $C_F=\frac{N_c^2-1}{2\,N_c}$, and $T_F=\frac12$.

\subsection{Unpolarized lepton, Transversely polarized nucleon}\label{appendix:polarized}
The hard scattering coefficients that appear in the transverse nucleon spin structure functions $F_{UT}^{\sin\phi_s}$ in Eq.~\eqref{eq:F_UT@NLO} 
are presented below. For the $q\to qg$-fragmentation channel,  we separate the coefficient function in terms of the color factors $C_F$ and $N_c$,
\begin{equation}
    \mathcal{C}_{UT}^{q\to qg}(v,w,\zeta,\chi_Q) = C_F\,\mathcal{C}_{UT,C_F}^{q\to qg}(v,w,\zeta,\chi_Q) + N_c\,\mathcal{C}_{UT,N_c}^{q\to qg}(v,w,\zeta,\chi_Q)\,.\label{eq:CUTcolorsplit}
\end{equation}
Then, the part of the coefficient function $\mathcal{C}_{UT}^{q\to qg}$ proportional to $C_F$ reads,
\begin{eqnarray}
    \mathcal{C}_{UT,C_F}^{q\to qg}(v,w,\zeta,\chi_Q) & = & \delta(1-v)\,\delta(1-w)\,\left[\frac{2\,\ln\zeta-\ln^2\zeta}{2\,(1-\zeta)}+\frac{(2-2\,\zeta-\ln\zeta)\,\ln\chi_Q}{1-\zeta}-\frac{7}{2}\right]\nn\\
    &+&\,\delta(1-v)\,2\,w^2\,\left[\left(\frac{\ln(1-w)}{1-w}\right)_++\frac{1+\ln\chi_Q-\ln w}{(1-w)_+}\right]\nn\\
    &+&\,\delta(1-w)\,\left[1-\frac{1-\zeta-v\,(3-\zeta)+2\,v^2\,(1-\zeta)}{\zeta}\left\{\left(\frac{\ln(1-v)}{1-v}\right)_++\frac{1+\ln\chi_Q+\ln v}{(1-v)_+}\right\}\right]\nn\\
    &+&w^2\,\frac{(1-v)^2\,\zeta^2-w\,\zeta\left[\zeta-v\,(3+\zeta)+3\,v^2-2\,v^3\,(1-\zeta)\right]-w^2\,(1-v)(1-2v)\,(1-\zeta)}{\zeta\,\left[(1-v)\,\zeta+w\,(1-\zeta)\right]\,(1-v)_+\,(1-w)_+}\,,\label{eq:CUTCF}
\end{eqnarray}
while the part proportional to $N_c$ takes the form,
\begin{eqnarray}
    \mathcal{C}_{UT,N_c}^{q\to qg}(v,w,\zeta,\chi_Q) & = & \delta(1-v)\,\delta(1-w)\,\left[-\frac{\ln(1-\zeta)}{2\,\zeta}-\frac{(2-2\,\ln\chi_Q-\ln\zeta)\,\ln\zeta}{4\,(1-\zeta)}\right]\nn\\
    &&\hspace{-3cm}+\delta(1-w)\,\left[\frac{1+\zeta-v\,(2-\zeta+\zeta^2)}{2\,\zeta\,(1-\zeta)\,(1-v\,\zeta)}\left(\ln\chi_Q+\ln(1-v)+\ln v\right)+\frac{1+\zeta-2\,v\,(1+\zeta^2)+2\,v^2\,\zeta^2}{2\,\zeta\,(1-\zeta)\,(1-v\,\zeta)}\right]\nn\\
    &&\hspace{-2cm}+\frac{w^2\,\left(2\,(1-v)\zeta^2+3\,(1-v)\,w\,\zeta\,(1-\zeta)+(1-2v)\,w^2\,(1-\zeta)^2\right)}{2\,\zeta\,(1-\zeta)\,[(1-v)\,\zeta+w\,(1-\zeta)]\,(1-w)_+}\,.\label{eq:CUTNc}
\end{eqnarray}
For the partonic $q\to\bar{q}q$ channel, we obtain the following explicit result:
\begin{eqnarray}
    \mathcal{C}_{UT}^{q\to \bar{q}q}(v,w,\zeta,\chi_Q) & = & -\frac{1}{2\,N_c}\Bigg\{-\frac{w^3\,(1-v)\,(1-2v)}{(1-\zeta)\,v\,[(1-v)\,\zeta+(1-\zeta)\,w]}\nn\\
    &&+\delta(1-w)\,\left[\frac{(1-v)(1-2v)\left(\ln\chi_Q+\ln(1-v)+\ln v+1\right)}{\zeta\,[1-v\,(1-\zeta)]}+\frac{v}{[1-v\,(1-\zeta)]}\right]\nn\\
    &&+\frac{w^3\,(1-v)\,(1-2v)}{\zeta\,[(1-v)(1-\zeta)+w\,\zeta]\,(1-w)_+}\Bigg\}\,.\label{eq:CUTq2qq}
\end{eqnarray}
Note that the coefficient function $ \mathcal{C}_{UT}^{q\to \bar{q}q}$ is $\mathcal{O}(1/N_c^2)$-suppressed in comparison with the coefficient function $ \mathcal{C}_{UT}^{q\to qg}$. In the large $N_c$-limit, this channel becomes irrelevant.

\subsection{Longitudinally polarized lepton, Transversely polarized nucleon\label{app:LT}}

The hard scattering coefficients that appear in the transverse nucleon spin structure functions $F_{LT}^{\cos\phi_s}$ in Eq.~\eqref{eq:F_LT@NLO} 
are presented below. For the $q\to qg$-fragmentation channel,  we separate the coefficient function in terms of the color factors $C_F$ and $N_c$,
\begin{equation}
    \mathcal{C}_{LT}^{q\to qg}(v,w,\zeta,\chi_Q) = C_F\,\mathcal{C}_{LT,C_F}^{q\to qg}(v,w,\zeta,\chi_Q) + N_c\,\mathcal{C}_{LT,N_c}^{q\to qg}(v,w,\zeta,\chi_Q)\,.\label{eq:CLTcolorsplit}
\end{equation}
Then, the part of the coefficient function $\mathcal{C}_{LT}^{q\to qg}$ proportional to $C_F$ reads:
\begin{eqnarray}
    \mathcal{C}_{LT,C_F}^{q\to qg}(v,w,\zeta,\chi_Q) & = & \delta(1-v)\,\delta(1-w)\,\left[\frac{2\,\ln\zeta-\ln^2\zeta}{2\,(1-\zeta)}+\frac{(2-2\,\zeta-\ln\zeta)\,\ln\chi_Q}{1-\zeta}-\frac{7}{2}\right]\nn\\
    &+&\,\delta(1-v)\,2\,w^2\,\left[\left(\frac{\ln(1-w)}{1-w}\right)_++\frac{1+\ln\chi_Q-\ln w}{(1-w)_+}\right]\nn\\
    &+&\,\delta(1-w)\,\left[-1+\frac{1-\zeta-v\,(1-3\zeta)}{\zeta}\left\{\left(\frac{\ln(1-v)}{1-v}\right)_++\frac{1+\ln\chi_Q+\ln v}{(1-v)_+}\right\}\right]\nn\\
    &+&w^2\,\frac{(1-v)\,w-\zeta+3\,v\,\zeta}{\zeta\,(1-v)_+\,(1-w)_+}\,,\label{eq:CLTCF}
\end{eqnarray}
while the part proportional to $N_c$ takes the form,
\begin{eqnarray}
    \mathcal{C}_{LT,N_c}^{q\to qg}(v,w,\zeta,\chi_Q) & = & \delta(1-v)\,\delta(1-w)\,\left[-\frac{\ln(1-\zeta)}{2\,\zeta}-\frac{(2-2\,\ln\chi_Q-\ln\zeta)\,\ln\zeta}{4\,(1-\zeta)}\right]\nn\\
    &&\hspace{0cm}+\delta(1-w)\,\left[-\frac{1}{2\,\zeta}\left(\ln\chi_Q+\ln(1-v)+\ln v\right)-\frac{1-2\,v\,\zeta}{2\,\zeta\,(1-v\,\zeta)}\right]-\frac{w^3}{2\,\zeta\,(1-w)_+}\,.\label{eq:CLTNc}
\end{eqnarray}
For the partonic $q\to\bar{q}q$ channel, we obtain the following explicit result:
\begin{eqnarray}
    \mathcal{C}_{LT}^{q\to \bar{q}q}(v,w,\zeta,\chi_Q) & = & -\frac{1}{2\,N_c}\Bigg\{-\frac{w^3\,(1-v)}{(1-\zeta)\,v\,[(1-v)\,\zeta+(1-\zeta)\,w]}\nn\\
    &&+\delta(1-w)\,\left[\frac{(1-v)\left(\ln\chi_Q+\ln(1-v)+\ln v+1\right)}{\zeta\,[1-v\,(1-\zeta)]}+\frac{v}{[1-v\,(1-\zeta)]}\right]\nn\\
    &&+\frac{w^3\,(1-v)}{\zeta\,[(1-v)(1-\zeta)+w\,\zeta]\,(1-w)_+}\Bigg\}\,.\label{eq:CLTq2qq}
\end{eqnarray}
Note that, again, the coefficient function $ \mathcal{C}_{LT}^{q\to \bar{q}q}$ is $\mathcal{O}(1/N_c^2)$-suppressed in comparison with the coefficient function $ \mathcal{C}_{LT}^{q\to qg}$. In the large $N_c$-limit, this channel becomes irrelevant.

\clearpage

\bibliography{Referenzen}

\end{document}